\documentclass[12pt,english,preprint]{article}
\usepackage{graphicx}
\usepackage{amsfonts}
\usepackage[T1]{fontenc}
\usepackage[latin9]{inputenc}
\usepackage{geometry}
\usepackage{amstext}
\usepackage{amssymb}
\usepackage{setspace}
\usepackage{esint}
\def\eq#1{{eq.~(\ref{#1})}}
\def\Eq#1{{Eq.~(\ref{#1})}}

\def\pim{{\bf\pi}}

\def\Tr{\mbox{Tr}\,}
\def\tr{\mbox{tr}\,}

\def\hbar{\hspace{0pt}\raisebox{1pt}{$-$} \hspace{-7pt} h}

\def\5{\overline 5}

\newcommand{\be}{\begin{equation}}
\newcommand{\ee}{\end{equation}}
\newcommand{\bea}{\begin{eqnarray}}
\newcommand{\eea}{\end{eqnarray}}
\newcommand{\nn}{\nonumber}

\usepackage{slashed}

\begin{document}
%%%%%%%%%%%%%%%%%%%%%%%%%%%%%%%%%%%%%%%%%%%%%%%%%%%%%%%%%%%  FRONT PAGE

\title{
\vspace{-1cm}
\bf
Computing the Effective Action\\
with the Functional Renormalization Group
\vspace{1cm}
}

\vspace{3cm}

\author{
Alessandro Codello$^{a}$,
Roberto Percacci$^{b}$,
Les\l{}aw Rachwa\l{}$^{c}$ and
Alberto Tonero$^{d}$\\\\
$^{a}$\emph{CP3-Origins \& the Danish IAS University of Southern Denmark,}\\
\emph{Campusvej 55, DK-5230 Odense M, Denmark}\\
$^{b}$\emph{SISSA, via Bonomea 265, 34136 Trieste, Italy}\\
\emph{and INFN, sezione di Trieste, Italy}\\
$^{c}$\emph{Department of Physics \& Center for Field Theory and Particle Physics,}\\
\emph{Fudan University, 200433 Shanghai, China}\\
$^{d}$\emph{ICTP-SAIFR \& IFT, Rua Dr.\ Bento Teobaldo Ferraz 271,}\\
\emph{01140-070 S\~ao Paulo, Brazil} \\
\vspace{1cm}
}

\date{}
\maketitle

\begin{abstract}
The ``exact'' or ``functional'' renormalization group equation describes the renormalization group flow
of the effective average action $\Gamma_k$. %, as a function of a cutoff $k$. 
The ordinary effective action $\Gamma_0$ can be obtained by integrating the flow equation from an ultraviolet scale $k=\Lambda$ down to $k=0$.
We give several examples of such calculations at one-loop, both in renormalizable and in effective field theories. 
We use the results of Barvinsky, Vilkovisky and Avramidi on the non-local heat kernel coefficients to reproduce the four point scattering amplitude in the case of a real scalar field theory with quartic potential
and in the case of the pion chiral lagrangian.
In the case of gauge theories, we reproduce the vacuum polarization of QED and of Yang-Mills theory.
We also compute the two point functions for scalars and gravitons in the effective field theory of scalar fields minimally coupled to gravity.
\end{abstract}

\vskip1.5em

\newpage

\tableofcontents{}

%\newpage

%%%%%%%%%%%
%%%%%%%%%%%
\section{Introduction}
%%%%%%%%%%%
%%%%%%%%%%%

The functional renormalization group (FRG) is a way of studying the flow
of infinitely many couplings as functions of an externally imposed cutoff.
The idea originates from Wilson's understanding of the renormalization group (RG)
as the change in the action that is necessary to obtain the same partition function
when the ultraviolet (UV) cutoff is lowered \cite{wilson}.
Early implementations of this idea were based on discrete RG transformations,
but soon there appeared equations describing the change of the action
under continuous shifts of the cutoff.
The first such equation was the Wegner--Houghton equation \cite{wegnerhoughton},
which has been widely used to study statistical models and, 
in a particle physics context, to put bounds on the Higgs mass \cite{hasenfratz}.
Another related equation that has been used originally to gain new insights 
in the renormalizability of $\phi^4$ theory is the Polchinski equation \cite{polchinski}.
In particle physics one is usually more interested in the effective action (EA) than
in the partition function, so one may anticipate that an equation describing the
flow of the generator of 1PI Green functions may be of even greater use.
For this purpose, the convenient functional to use is the effective average action (EAA) $\Gamma_k$.
It is defined in the same way as the ordinary effective action, with the following two
modifications: first, one adds to the bare action a cutoff term characterized by a cutoff scale $k$, of the form
\begin{equation}
\Delta S_{k}[\phi]=\frac{1}{2}\int d^{d}q\,\phi(-q)\, R_{k}(q^2)\,\phi(q)
\label{cutoff}
\end{equation}
and second, after performing the Legendre transform one subtracts the same term.
For general reviews see e.g. \cite{morrisrev,bb,btw,delamotte}.
The effect of this term is to suppress the propagation of low momentum modes
leaving the vertices unchanged.
The cutoff kernel $R_k(z)$ is required
to go to zero fast when its argument $z$ (which in flat spacetime applications
can be thought of as momentum squared) is greater than the cutoff scale $k^2$.
In typical application this decay could be a polynomial of sufficiently high degree
or an exponential. 
The cutoff kernel is also required to tend to zero (for all $z$) when $k\to0$.
This implies that when $k\to0$ the EAA reduces to the ordinary effective action.

The $k$--dependence of the EAA is described by the Wetterich equation
\cite{wetterich,morris1}
\begin{equation}
\label{erge}
\partial_t\Gamma_k=
\frac{1}{2}\mathrm{STr}\left(\frac{\delta^{2}\Gamma_k}
{\delta\phi\delta\phi}+R_{k}\right)^{-1}\partial_t R_k\ , 
\end{equation}
where $t=\log (k/k_0)$, $k_0$ is an arbitrary reference scale
and the supertrace in the r.h.s. stands (in flat spacetime) for an integration
over coordinate and momentum space and a trace over any representation
of internal and spacetime symmetries that the fields may carry.
Due to the fast fall-off of the cutoff kernel, also
the function $\partial_t R_k$, which appears inside the trace
in the r.h.s. of (\ref{erge}), decays fast for large $z$.
This makes the trace in the r.h.s. of (\ref{erge}) convergent.

The functional renormalization group equation (FRGE) has been widely used in studies of the infrared (IR) properties
of statistical and particle physics models, in particular of phase transitions 
and critical phenomena.
It has also been used to study the ultraviolet behavior of gravity,
in particular to establish the existence of a nontrivial fixed point which
may be used to define a continuum limit
\cite{rn,percaccirev,litimrev}.
Here we would like to discuss some examples taken mostly from particle physics 
where the Wetterich equation is used instead as a tool to compute the effective action.

The basic idea is as follows. 
Assume that $\Gamma_k$ is the most general functional of the given fields
which is invariant under the assumed symmetries of the system.
In many application it is justified to assume that it is a semi-local functional \cite{semilocal},
meaning that it admits an expansion into infinitely many local terms
constructed with the fields and their derivatives of arbitrary degree.
We call ``theory space'' the space of these functionals.
\Eq{erge} defines a vector-field on this space
whose integral lines are the RG trajectories.
We can now fix an arbitrary initial point in theory space and identify it as
the ``bare'' action of the theory at some UV scale $\Lambda$.
Typically one will choose this bare action to be local and simple,
but this is not essential.
One can integrate the RG flow in the direction of decreasing $t$
and the IR endpoint of the flow for $t\to-\infty$ represents the effective action.
The couplings in the effective action can be interpreted as renormalized couplings,
and the integral of their beta functions from $k=0$ to $k=\Lambda$
is the relation between bare and renormalized couplings.

One can also ask what would happen if we tried to take the limit $\Lambda\to\infty$.
This is equivalent to solving the FRGE in the direction of increasing $t$
with the same initial condition.
(Since the initial condition at the original value of $\Lambda$ is kept fixed,
also the effective action will remain fixed, so this is very similar to
the Wilsonian procedure of changing the action at the UV cutoff ``keeping the physics fixed''.) 
There is a variety of possible behaviors. If some coupling blows up at finite $t$
(a Landau pole), the RG flow stops there and one has to interpret the theory as an effective
field theory with an upper limit to its validity.
On the other hand if the trajectory ends at a fixed point,
one may expect all physical observables to be well behaved.
In this case the theory is UV complete.
The main point is that by integrating the flow towards the UV one can study the
UV divergences of the theory and argue about its UV completeness.

Below we will calculate the r.h.s. of the Wetterich equation for several theories,
and then integrate the flow down to $k=0$ to obtain the effective action.
Of course, given that the effective action of any nontrivial theory is
infinitely complicated, we can only obtain partial information about it,
and then only in certain approximations.
Here we will exploit the great flexibility of the FRGE with regards to approximation schemes.
In typical previous applications of the FRGE, for example in the study of the
Wilson--Fisher fixed point, it is often enough to retain only the zero--momentum part of
the effective action, but it is important to retain the full field dependence.
In particle physics one usually considers the scattering of a few particles
at the time and the full field dependence is not needed.
On the other hand, one is interested in the full momentum dependence.
Clearly, a different type of approximation is needed.

In what follows, unless otherwise stated, 
we will calculate the r.h.s. of the flow equation
keeping $\Gamma_k$ fixed at its ultraviolet form $\Gamma_\Lambda=S$.
\footnote{There is some danger in this terminology:
the action that is used in the functional integral to define the theory actually differs from $\Gamma_\Lambda$
by a determinant. We hope this will not generate confusion.}
In perturbation theory this is equivalent to working at one loop.
The one-loop EAA is given by
\be
\Gamma^{(1)}_{k}=S+\frac{1}{2}\textrm{Tr}\log
\left(\frac{\delta^{2}S}{\delta\phi\delta\phi}+R_{k}\right)
\ee
and satisfies the equation
\begin{equation}
\label{oneloopERGE}
\partial_t\Gamma_k^{(1)}=\frac{1}{2}\mathrm{Tr}\left(\frac{\delta^{2}S}
{\delta\phi\delta\phi}+R_{k}\right)^{-1}\partial_t R_k\ .
\end{equation}
We will use known results on the non-local heat kernel to compute the trace on the r.h.s.
and in this way obtain the flow of the non-local part of the EAA. 
A similar calculation in the full flow equation is beyond currently available techniques.
Integrating the flow we will derive the non-local, finite parts of the
effective action containing up to two powers of the field strength.
(By ``field strength'' we mean here in general the curvature of the metric or of the
Yang-mills fields or the values of the scalar condensates).
Such terms can be used to describe several physical phenomena,
such as self-energies and, in some cases, scattering amplitudes.
In each of the cases that we shall consider one can find some justification
for the approximations made, as we shall discuss below.

We now give a brief overview  of the main results and of the content of the subsequent sections.
In section II we review the mathematical results for the 
non--local heat kernel expansion of a function of a Laplace--type operator of \eq{operator}.
The operator will generally depend on background fields
such as metric, Yang-Mills field (if it acts on fields carrying nonzero charges 
of the gauge group) or scalar condensates.
The trace of the heat kernel of the operator admits a well know asymptotic expansion
whose coefficients are integrals of powers of the field strengths.
The non-local heat kernel expansion is a sum of infinitely many such terms,
containing a fixed number of field strengths but arbitrary powers of derivatives.
It can thus be viewed as a vertex expansion of the heat kernel.
One can expand the trace of any operator, and hence also the trace on the r.h.s.
of \eq{erge}, as a sum of these non-local expressions,
with coefficients that depend on the particular function that is under the trace.
There are certain ambiguities in these calculations:
one may choose to regard the r.h.s. as a function of different operators,
and one has the freedom of choosing different cutoff functions $R_k$.
We shall see, however, that physical results are independent of these choices.

As a warmup in section III we will begin by using this technique to calculate the EA of a scalar field.
%the self-energy of a scalar field.
We will see that the integration of the FRGE yields the familiar relations
between the bare and renormalized couplings and that in the limit $\Lambda\to\infty$
there are only three divergences.
Integration of the flow equation down to $k=0$ yields an EA
that encodes the familiar formula for the one-loop scattering amplitude.

In section IV we compute the EA for photons in QED
which is obtained by integrating out the fermion fields.
We reproduce the known vacuum polarization effects, 
and, within an expansion in inverse powers of the electron mass,
the four-photon interactions described by the Euler-Heisenberg Lagrangian.
In section V we calculate the vacuum polarization 
effects in Yang-Mills theory.
In this case, unlike all other cases considered in this paper, due to IR divergences it is not possible
to integrate the flow equation down to $k=0$.
We thus have to restrict our attention to a finite range of momenta for which the theory
remains in a neighborhood of its asymptotically free UV fixed point.

The remaining two sections are devoted to examples of effective field theories (EFTs).
In section VI we consider the chiral nonlinear sigma model,
which describes the low energy interactions of pions
(and also, in a different interpretation, the low energy scattering of longitudinal $W$ bosons).
As expected, in this case we find divergences that are not of the same form of the original action.
The effective action is organized as an expansion in powers of $p/ F_\pi$,
where $p$ is momentum and $F_\pi$ is the pion decay constant.
%The one-loop approximation in this case is justified by an 
We compute in the one-loop approximation the four point function and we show that
it reproduces the well known result of Gasser and Leutwyler \cite{gl}.

Finally in section VII we consider the theory of a scalar field coupled to dynamical gravity.
We compute the FRGE keeping terms with two field strengths
but all powers of momentum.
This calculation is justified by an expansion in powers of $p/M_{\rm Planck}$.
In this case we obtain for the first time unambiguous covariant formulae 
for the logarithmic terms in the EA.

%%%%%%%%%%%%%%%%%%%%%%%
%%%%%%%%%%%%%%%%%%%%%%%
\section{The non--local heat kernel expansion}
%%%%%%%%%%%%%%%%%%%%%%%
%%%%%%%%%%%%%%%%%%%%%%%

The r.h.s. of equation (\ref{erge}) is the trace of a function of an operator $\Delta$.
In the simplest cases this operator is a second-order Laplace-type operator.
In the presence of general background fields (gravity, Yang-Mills fields)
this operator will be a covariant Laplacian,
related to the inverse propagator of the theory in question.
In general it will have the form 
\be
\label{operator}
\Delta=-D^2\mathbf{1}+\mathbf{U}\ ,
\ee
where $D$ is a covariant derivative with respect to all the background fields
and $\mathbf{U}$ is a non-derivative part that is a matrix in the appropriate
representations of all the symmetry groups that are carried by the fields
(it thus carries both internal and spacetime indices).

Before discussing any physical application, we outline here the heat kernel method
we employ in the calculation of the trace.
The typical expression that we need to trace is
\be
\label{function}
h_k(\Delta,\omega)=\frac{\partial_t R_k(\Delta)}{\Delta+\omega+R_k(\Delta)}\ .
\ee
Sometimes one has an additional term $-\eta R_k(\Delta)$ in the numerator,
where $\eta$ is called `anomalous dimension'. This term can be neglected in one loop calculations.
The typical form of the Wetterich equation is then
\be
\label{wetterich}
\partial_t\Gamma_k=
\frac{1}{2}\Tr h_k(\Delta,\omega)\,.
\ee
Let us introduce the Laplace transform $\tilde h_k(s,\omega)$ by
\be
h_k(\Delta,\omega)=\int_0^\infty ds\, \tilde h_k(s,\omega) e^{-s\Delta} \ .
\label{laplace}
\ee
If we insert \eq{laplace} in the r.h.s. of \eq{wetterich}, by linearity the trace goes through the integral and we remain with
\be
\label{elsa}
\partial_t\Gamma_k=
\frac{1}{2}\int_0^\infty ds\, \tilde h_k(s,\omega) \Tr e^{-s\Delta} \ .
\ee
One can now use the asymptotic expansion for the trace of the heat kernel
\be
\label{hk}
\Tr e^{-s\Delta}=\frac{1}{(4\pi s)^{d/2}}\int d^dx\sqrt{g}\,
\tr \left[\mathbf{b}_0(\Delta)+\mathbf{b}_2(\Delta) s+\mathbf{b}_4(\Delta)s^2+\ldots+\mathbf{b}_d(\Delta)s^{d/2}+\ldots\right]\,,
\ee
whose first three coefficients are:
\bea
\mathbf{b}_0(\Delta) & = & \mathbf{1}
\nn\\
\mathbf{b}_2(\Delta) & = & \frac{R}{6}\mathbf{1}-\mathbf{U}
\nn\\
\mathbf{b}_4(\Delta) & = & \frac{1}{2}\mathbf{U}^2+\frac{1}{6}D^2\mathbf{U}+\frac{1}{12}\mathbf{\Omega}_{\mu\nu}\mathbf{\Omega}^{\mu\nu}-\frac{R}{6}\mathbf{U}
\nn\\
&& +\mathbf{1}\left(\frac{1}{180}R_{\mu\nu\alpha\beta}^2-\frac{1}{180}R_{\mu\nu}^2+\frac{1}{72}R^2-\frac{1}{30}D^2R\right)
\label{hkcoefficient}
\eea
where the space-time curvatures are constructed using the Levi-Civita connection and $\mathbf{\Omega}_{\mu\nu}=[D_\mu,D_\nu]$ is the 
field strength tensor.
The first $d/2$ terms in \eq{hk} come with an overall negative power of $s$, while all subsequent terms have positive powers.
When we insert this expansion in \eq{elsa} we can write
\bea
\partial_t\Gamma_k=
\frac{1}{2}\frac{1}{(4\pi)^{d/2}}\int d^d x\sqrt{g}\,\tr
\bigg\lbrace\mathbf{b}_0(\Delta)Q_{\frac{d}{2}}[h_k]
+\mathbf{b}_2(\Delta)Q_{\frac{d}{2}-1}[h_k]  \qquad\qquad\qquad\nn \\ 
\qquad\qquad\qquad+ \, \mathbf{b}_4(\Delta)Q_{\frac{d}{2}-2}[h_k]
+\dots +\mathbf{b}_d(\Delta)Q_0[h_k]+\ldots \bigg\rbrace
\label{elsa2}
\eea
where the ``$Q$--functionals'' are defined by
\be 
\label{q1}
Q_n[f]=\int_0^\infty ds s^{-n}\tilde f(s)\ .
\ee
For $n$ a positive integer one can use the definition of the
Gamma function to rewrite (\ref{q1}) as a Mellin transform
\be
Q_{n}[f]=\frac{1}{\Gamma(n)}\int_{0}^{\infty}dw\, w^{n-1}f(w)\,,
\label{Qnpos}
\ee
while for $m$ a positive integer or $m=0$
\begin{equation}
Q_{-m}[f] = (-1)^{m}f^{(m)}(0)\ .
\label{Qnneg}
\end{equation}
This expansion is useful to study the UV divergences, which are always given
by local expressions. In particular, one finds that the first $d$ terms 
in the expansion (\ref{elsa2})
give rise to divergences in the effective action.

In order to calculate non-local, finite parts of the effective action 
we need a more sophisticated version of the heat kernel
expansion which includes an infinite number of heat kernel coefficients.
This expansion has been developed in \cite{bv1,bv2,avra1,avra2}
and retains the infinite number of heat kernel coefficients in the
form of non-local ``structure functions'' or ``form factors''.
For an alternative derivation see \cite{Codello:2012kq}.
Keeping terms up to second order in the fields strengths,
the non--local heat kernel expansion reads as follows:
\begin{eqnarray}
\Tr e^{-s\Delta} &=&  \frac{1}{(4\pi s)^{d/2}}
\int d^{d}x\sqrt{g}\,\tr\bigg\lbrace \mathbf{1}-s \mathbf{U}+s\mathbf{1}\frac{R}{6}
+ s^{2}\left[\mathbf{1}R_{\mu\nu}f_{Ric}(-sD^2)R^{\mu\nu}
\right.\nonumber \\
&&\left.
\qquad\qquad%\qquad
+ \mathbf{1}R\, f_{R}(-sD^2)R
+ \, R\,f_{RU}(-sD^2)\mathbf{U}
+ \mathbf{U}f_{U}(-sD^2)\mathbf{U}
\right.\nonumber \\
&&\left.
\qquad\qquad
+ \, \mathbf{\Omega}_{\mu\nu}f_{\Omega}(-sD^2)\mathbf{\Omega}^{\mu\nu}\right]
+\ldots\bigg\rbrace \,.
\label{nlhk}
\end{eqnarray}
The structure functions in \eq{nlhk} are given by
\begin{eqnarray}
f_{Ric}(x) & = & \frac{1}{6x}+\frac{1}{x^{2}}\left[f(x)-1\right]\nn
\\
f_{R}(x) & = & \frac{1}{32}f(x)+\frac{1}{8x}f(x)-\frac{7}{48x}
-\frac{1}{8x^{2}}\left[f(x)-1\right]\nn
\\
f_{RU}(x) & = & -\frac{1}{4}f(x)-\frac{1}{2x}\left[f(x)-1\right]\nn
\\
f_{U}(x) & = & \frac{1}{2}f(x)\nonumber 
\\
f_{\Omega}(x) & = & -\frac{1}{2x}\left[f(x)-1\right]\,,
\label{sf}
\end{eqnarray}
where the basic heat kernel structure function $f(x)$ is defined
in terms of the parametric integral
\begin{equation}
f(x)=\int_{0}^{1}d\xi\, e^{-x\xi(1-\xi)}\,.
\label{bsf}
\end{equation}

Using in \eq{sf} the Taylor expansion of the basic structure function $f(x)=1-\frac{x}{6}+\frac{x^{2}}{60}+O(x^{4})$,
we obtain the following ``short time'' expansion for the structure functions
\begin{eqnarray}
f_{Ric}(x) & = & \frac{1}{60}-\frac{x}{840}+\frac{x^{2}}{15120}+O(x^{4})\nonumber \\
\, f_{R}(x) & = & \frac{1}{120}-\frac{x}{336}+\frac{11x^{2}}{30240}+O(x^{4})\nonumber \\
f_{RU}(x) & = & -\frac{1}{6}+\frac{x}{30}-\frac{x^{2}}{280}+O(x^{4})\nonumber \\
f_{U}(x) & = & \frac{1}{2}-\frac{x}{12}+\frac{x^{2}}{120}+O(x^{4})\nonumber \\
f_{\Omega}(x) & = & \frac{1}{12}-\frac{x}{120}+\frac{x^{2}}{1680}+O(x^{4})\,.\label{sfexp}
\end{eqnarray}
If we insert \eq{sfexp} in \eq{nlhk}, the first term reproduces the
coefficients of the local heat kernel expansion discussed previously.
If we compare with \eq{hk} we see that not
all coefficients match exactly. This is because the local heat kernel
expansion is derived by calculating the un-integrated coefficients
while the non-local heat kernel expansion is derived by calculating
the integrated ones. So the coefficients derived by expanding the
structure functions \eq{sf} may differ from the local ones
\eq{hkcoefficient} by a total derivative or a boundary term. For example,
only two of the three possible curvature square invariants present
in \eq{hk} appear in \eq{nlhk}, the third one has been
eliminated using Bianchi's identities and discarding a boundary term.
For this reason also the total derivative terms in the coefficient
\textbf{$B_{4}(\Delta)$} are not present in the non-local expansion.
Thus, in general, a straightforward series expansion of the non-local
heat kernel structure functions will not reproduce exactly the same
heat kernel coefficients of the local expansion. See \cite{bv2}
for more details on this point. 

Inserting \eq{nlhk} in \eq{elsa} we obtain
\begin{eqnarray}
\partial_t\Gamma_k & = & 
\frac{1}{2}\frac{1}{(4\pi)^{d/2}}
\int d^{d}x\sqrt{g}\,\tr\Bigg\lbrace
\mathbf{1}\left[\int_0^\infty ds\, \tilde h_k(s,\omega)s^{-d/2}\right]
\nonumber \\
 &  & 
-\mathbf{U}\left[\int_0^\infty ds\, \tilde h_k(s,\omega)s^{-d/2+1}\right]
+\frac{R}{6}\mathbf{1}\left[\int_0^\infty ds\, \tilde h_k(s,\omega)s^{-d/2+1}\right]
\nonumber \\
 &  & 
+\mathbf{1}R_{\mu\nu}\left[\int_0^\infty ds\, \tilde h_k(s,\omega)s^{-d/2+2}f_{Ric}(sz) \right]R^{\mu\nu}
+\mathbf{1}R\left[\int_0^\infty ds\, \tilde h_k(s,\omega)s^{-d/2+2}f_{R}(sz) \right]R
\nonumber \\
 &  & 
+R\left[\int_0^\infty ds\, \tilde h_k(s,\omega)s^{-d/2+2}f_{RU}(sz) \right]\mathbf{U}
+ \mathbf{U}\left[\int_0^\infty ds\, \tilde h_k(s,\omega)s^{-d/2+2}f_{U}(sz) \right]\mathbf{U}
\nonumber \\
 &  & 
+ \mathbf{\Omega}_{\mu\nu} \left[ \int_0^\infty ds\, \tilde h_k(s,\omega)s^{-d/2+2}f_{\Omega}(sz) \right]\mathbf{\Omega}^{\mu\nu}
+O(\mathcal{R}^{3})\Bigg\rbrace
\nonumber\\
&=&
\frac{1}{2}\frac{1}{(4\pi)^{d/2}}
\int d^{d}x\sqrt{g}\,\Bigg\lbrace
Q_{\frac{d}{2}}[h_k]\,\tr\!\mathbf{1}
+Q_{\frac{d}{2}-1}[h_k]\,\tr\!\!\left(\frac{R}{6}\mathbf{1}-\mathbf{U}\right)
+R_{\mu\nu}g_{Ric}R^{\mu\nu}\,\tr\!\mathbf{1}
\nonumber \\
 &  & 
+\,R\,g_R R\,\tr\!\mathbf{1}
+R\,g_{RU}\,\tr\!\mathbf{U}
+\tr\! \Big( \mathbf{U}g_U\mathbf{U} \Big)
+ \tr \!\Big( \mathbf{\Omega}_{\mu\nu}\,g_\Omega\, \mathbf{\Omega}^{\mu\nu} \Big)
+\ldots\Bigg\rbrace \ .
\label{adam}
\end{eqnarray}
Here and in the following $z=-D^2$.
The first three terms are local and have been rewritten in terms of
$Q$--functionals as in the first terms of \eq{elsa2}.
In the remaining ones we have defined
\be\label{giAfunctions}
g_A=g_A( z, \omega,k)= \int_0^\infty ds\, \tilde h_k(s,\omega)s^{-d/2+2}f_A(sz)\,,
\ee
for $A=\{Ric,R,U,RU,U,\Omega\}$.
From the definition of the Laplace transform we see that shifting the argument
of $h_k$ by $a$ is the same as multiplying the Laplace transform by $e^{-sa}$.
Then:
\be\label{shiftq}
\int_0^\infty ds\,s^{-n}e^{-sa}\tilde h_k(s,\omega)=Q_n[h_{k,\omega}^a]\,,
\ee
where
\be
h_{k,\omega}^a(z)\equiv h_k(z+a,\omega)\ ;\qquad
h_{k,\omega}(z)\equiv h_k(z,\omega)\ .
\ee
We can use this to write the functions $g_A$ in terms of 
$Q$--functionals of shifted arguments:
\begin{eqnarray}
g_{Ric}(z,\omega,k)\!\!\!&=&\!\!\!\!
\frac{1}{6z}Q_{\frac{d}{2}-1}[h_{k,\omega}]-
\frac{1}{z^2}Q_{\frac{d}{2}}[h_{k,\omega}]
+\frac{1}{z^2}\int_0^1 d\xi\, Q_{\frac{d}{2}}[h_{k,\omega}^{z\xi(1-\xi)}]
\nonumber\\
g_R(z,\omega,k)\!\!\!&=&\!\!\!\!
-\frac{7}{48z}Q_{\frac{d}{2}-1}[h_{k,\omega}]
+\frac{1}{8z^2}Q_{\frac{d}{2}}[h_{k,\omega}]
+\frac{1}{32}\int_0^1 d\xi\, Q_{\frac{d}{2}-2}[h_{k,\omega}^{z\xi(1-\xi)}]
\nonumber\\
&&
+\frac{1}{8z}\int_0^1 d\xi\, Q_{\frac{d}{2}-1}[h_{k,\omega}^{z\xi(1-\xi)}]
-\frac{1}{8z^2}\int_0^1 d\xi\, Q_{\frac{d}{2}}[h_{k,\omega}^{z\xi(1-\xi)}]
\nonumber\\
g_{RU}(z,\omega,k)\!\!\!&=&\!\!\!
\frac{1}{2z}Q_{\frac{d}{2}-1}[h_{k,\omega}]
-\frac{1}{4}\int_0^1\!\! d\xi\, Q_{\frac{d}{2}-2}[h_{k,\omega}^{z\xi(1-\xi)}]
-\frac{1}{2z}\int_0^1 d\xi\, Q_{\frac{d}{2}-1}[h_{k,\omega}^{z\xi(1-\xi)}] 
\nonumber\\
g_U(z,\omega,k)\!\!\!&=&\!\!\!
\frac{1}{2}\int_0^1 d\xi\, Q_{\frac{d}{2}-2}[h_{k,\omega}^{z\xi(1-\xi)}]
\nonumber\\
g_\Omega(z,\omega,k)\!\!\!&=&\!\!\!
\frac{1}{2z}Q_{\frac{d}{2}-1}[h_{k,\omega}]
-\frac{1}{2z}\int_0^1 d\xi\, Q_{\frac{d}{2}-1}[h_{k,\omega}^{z\xi(1-\xi)}]\,.
\label{adam2}
\end{eqnarray}
These formulae can be made more explicit by choosing a specific cutoff kernel.
We will use the piecewise linear or ``optimized'' cutoff \cite{optimized}
\be
\label{optimized}
R_k(z)=(k^2-z)\theta(k^2-z)=k^2(1-\tilde z)\theta(1-\tilde z)\ ,
\ee
where $\tilde z=z/k^2$.
It has the virtue that the $Q$--functionals can be evaluated in closed form.
In $d=4$ we will need the functionals $Q_2$, $Q_1$ and $Q_0$
for both unshifted and shifted argument.
The unshifted $Q$--functionals are
\be
\label{Qopt}
Q_n\left[h_k\right]={2k^{2n}\over \Gamma(n+1)}\frac{1}{1+\tilde\omega}\qquad{\rm for}\ n=0,1,2\ldots
\ee
where $\tilde\omega=\omega/k^2$.
The parametric integrals of the shifted functionals can be calculated using
\begin{equation}
\int_{0}^{1}d\xi\, Q_{0}[h_{k}^{z\xi(1-\xi)}]=\frac{2k^2}{k^2+\omega}\left[1-\sqrt{1-\frac{4k^{2}}{z}}\theta(z-4k^{2})\right]
\label{HK_Q_9}
\end{equation}
\begin{equation}
\int_{0}^{1}d\xi\, Q_{1}[h_{k}^{z\xi(1-\xi)}]=\frac{2k^{4}}{k^2+\omega}\left[1-\frac{z}{6k^{2}}+\frac{z}{6k^{2}}\left(1-\frac{4k^{2}}{z}\right)^{\frac{3}{2}}\theta(z-4k^{2})\right]
\label{HK_Q_10}
\end{equation}
\begin{equation}
\int_{0}^{1}d\xi\, Q_{2}[h_{k}^{z\xi(1-\xi)}] =  \frac{k^{6}}{k^2+\omega}\left[1-\frac{z}{3k^{2}}+\frac{z^{2}}{30k^{4}}
-\frac{z^{2}}{30k^{4}}\left(1-\frac{4k^{2}}{z}\right)^{\frac{5}{2}}\theta(z-4k^{2})\right]\,.
\label{HK_Q_11}
\end{equation}
Plugging these expressions into \eq{adam2} we have that (in $d=4$) the functions $g_A$ depend only on $\tilde z$ and $\tilde \omega$. In this case it is convenient to define $g_A(\tilde z, \tilde \omega)\equiv g_A(z,\omega,k)$. These functions are explicitly given by
\begin{eqnarray}
g_{Ric}(\tilde z,\tilde \omega)\!&=&\!\!
\frac{1}{30}\frac{1}{1+\tilde\omega}\left[
1-\left(1-\frac{4}{\tilde z}\right)^{5/2}\theta(\tilde z-4)
\right] 
\label{eve1}
\\
g_R(\tilde z,\tilde \omega)\!&=&\!
\frac{1}{1+\tilde\omega}\Bigg[
\frac{1}{60}
-\frac{1}{16}\left(1-\frac{4}{\tilde z}\right)^{1/2}\theta(\tilde z-4)
\nonumber\\
&&
+\frac{1}{24}\left(1-\frac{4}{\tilde z}\right)^{3/2}\theta(\tilde z-4)
+\frac{1}{240}\left(1-\frac{4}{\tilde z}\right)^{5/2}\theta(\tilde z-4)
\Bigg]
\label{eve2}
\\
g_{RU}(\tilde z,\tilde \omega)\!&=&\!
\frac{1}{1+\tilde\omega}\left[
-\frac{1}{3}+\frac{1}{2}\sqrt{1-\frac{4}{\tilde z}}\theta(\tilde z-4)-\frac{1}{6}\left(1-\frac{4}{\tilde z}\right)^{\frac{3}{2}}\theta(\tilde z-4)
\right]
\label{eve3}
\\
g_U(\tilde z,\tilde \omega)\!&=&\!
\frac{1}{1+\tilde\omega}\left[
1-\left(1-\frac{4}{\tilde z}\right)^{1/2}\theta(\tilde z-4)
\right]
\label{eve4}
\\
g_\Omega(\tilde z,\tilde \omega)\!&=&\!
\frac{1}{6}\frac{1}{1+\tilde\omega}\left[
1-\left(1-\frac{4}{\tilde z}\right)^{3/2}\theta(\tilde z-4)
\right]\,.
\label{eve5}
\end{eqnarray}
With these relations we can now compute the functional traces on the rhs of the FRGE.

%%%%%%%%%%
%%%%%%%%%%
\section{Real scalar}
%%%%%%%%%%
%%%%%%%%%%

We begin considering Euclidean scalar theories
defined by the following bare action
\begin{equation}
S[\varphi]=\Gamma_\Lambda[\varphi]=\int d^dx\,\left\{
E_\Lambda+
\frac{1}{2}\partial_{\mu}\varphi\partial^{\mu}\varphi
+\frac{m_\Lambda^{2}}{2}\varphi^{2}
+\frac{\lambda_\Lambda}{4!}\varphi^4\right\}\,.
\label{eq:scalar}
\end{equation}
The field-independent term has been put in for later convenience
but it is unimportant as long as gravity can be neglected.
The restriction to a quartic potential is not dictated by arguments of renormalizability and to deal with arbitrary potential is not problematic in the context of the FRGE. In general, all the higher dimensional operators -- which are generated by the quartic interaction -- have an effect on the running of the quartic coupling itself. We ignore these terms because we are interested in reproducing the standard result for the one-loop four point amplitude.

Using \eq{eq:scalar}, the first step is to compute the Hessian entering in the one-loop RG flow of \eq{oneloopERGE}
\be 
\frac{\delta S}{\delta\varphi\delta\varphi}=-\Box+m_\Lambda^2+\frac{\lambda_\Lambda}{2}\varphi^2\,.
\ee
In order to properly account for threshold effects, it is convenient to choose the argument of the cutoff $R_k(\Delta)$ to be the operator $\Delta=-\Box+\frac{\lambda_\Lambda}{2}\varphi^2$.
Thus we have $\mathbf{U}=\frac{\lambda_\Lambda}{2}\varphi^2$.
In this way the function to be traced in the flow equation assumes the standard form of \eq{function} discussed in the previous section.
\Eq{elsa} reads
\be
\partial_{t}\Gamma_{k}[\varphi] 
 = \frac{1}{2}\Tr h_{k}\left(-\Box+\frac{\lambda_\Lambda}{2}\varphi^2,m_\Lambda^2\right)=\frac{1}{2}\int_{0}^{\infty}ds\,\tilde{h}_{k}(s,m_\Lambda^{2})\,\textrm{Tr}\, e^{-s\left(-\Box+\frac{\lambda_\Lambda}{2}\varphi^2\right)}\,.\label{scalarfe}
\ee
We evaluate the trace in \eq{scalarfe} using the non-local heat kernel expansion.
Setting $\mathbf{U}=\frac{\lambda_\Lambda}{2}\varphi^2$ in the general result of \eq{adam} gives:  
\be
\partial_{t}\Gamma_{k}[\varphi] = \frac{1}{2}\frac{1}{(4\pi)^{d/2}}\int d^dx\int_{0}^{\infty}ds\,\tilde{h}_{k}(s,m_{\Lambda}^{2})\, s^{-d/2}
\left[1-\frac{\lambda_\Lambda}{2}s\varphi^2
+\frac{\lambda_\Lambda^2}{4}s^{2}\varphi^2 f_U\left(sz\right)\varphi^2
\right]\,,
\ee
where $z=-\Box$.
Using \eq{q1} and \eq{adam2} we obtain the following form for the beta functional
\be
\partial_{t}\Gamma_{k}[\varphi]  =  \frac{1}{2}\frac{1}{(4\pi)^{d/2}}\int d^dx\left\{ Q_{\frac{d}{2}}[h_{k}]
-\frac{\lambda_\Lambda}{2}Q_{\frac{d}{2}-1}[h_{k}]\varphi^2
+\frac{\lambda_\Lambda^2}{8}\varphi^2\int_0^1 d\xi\, 
Q_{\frac{d}{2}-2}[h_k^{z \xi(1-\xi)}]
\varphi^2\right\} \,.
\label{scalarfrg}
\ee
At this point we make the following ansatz for the EAA entering in the l.h.s. of \eq{scalarfe}, bearing in mind that we are interested in terms up to fourth order in the scalar fields,
\begin{equation}
\Gamma_{k}[\varphi]=\int d^dx\,\left\{E_k+\frac{Z_{k}}{2}\partial_\mu\varphi\partial^\mu\varphi+\frac{m^2_k}{2}\varphi^2+\frac{1}{4!}\varphi^2 F_{k}(-\Box)\varphi^2\right\}\,,\label{ansatzscalargamma}
\end{equation}
with $F_{k}(0)=\lambda_k$. 
Plugging the ansatz for $\Gamma_k[\varphi]$ in \eq{scalarfrg} we read off the beta functions
\bea
\partial_t E_k & = & \frac{1}{2}\frac{1}{(4\pi)^{d/2}}Q_{\frac{d}{2}}\left[h_{k}\right]
\nonumber\\
\partial_{t}Z_{k} & = & 0
\nonumber\\
\partial_{t}m_{k}^{2} & = & -\frac{1}{2}\frac{\lambda_{\Lambda}}{(4\pi)^{d/2}}Q_{\frac{d}{2}-1}\left[h_{k}\right]
\label{betascalar}
\eea
and the flow of the structure function
\be
\partial_{t}F_{k}(z)=\frac{3}{2}\frac{\lambda_{\Lambda}^{2}}{(4\pi)^{d/2}}\int_{0}^{1}d\xi\, Q_{\frac{d}{2}-2}[h_{k}^{z\xi(1-\xi)}]\,.
\label{flowF}
\ee
Taking $z\to 0$ in the last equation we obtain the beta function for the self--interaction coupling constant
\begin{equation}
\partial_{t}\lambda_{k}  =  \frac{3}{2}\frac{\lambda_{\Lambda}^{2}}{(4\pi)^{d/2}}Q_{\frac{d}{2}-2}\left[h_{k}\right]\,.
\end{equation}
From now on we restrict ourselves to $d=4$, using \eq{Qopt} to evaluate the $Q$--functionals, we have that the beta functions for $m_{k}^{2}$ and $\lambda_{k}$ are
\begin{eqnarray}
\partial_{t}m_{k}^{2} & = & -\frac{\lambda_{\Lambda}}{(4\pi)^{2}}\frac{k^2}{k^2+{m}_{\Lambda}^{2}}\\
\partial_{t}\lambda_{k} & = & \frac{3\lambda_{\Lambda}^{2}}{(4\pi)^{2}}\frac{k^2}{k^2+{m}_{\Lambda}^{2}}\,.
\end{eqnarray}
One can now perform the RG improvement and make the substitutions $m_{\Lambda} \rightarrow m_{k}$, $\lambda_{\Lambda} \rightarrow \lambda_{k}$ in the r.h.s. of the beta functions and show, for example,
that the theory (\ref{eq:scalar}) in trivial. We will not repeat this discussion here, since we are interested to show how to compute the four point amplitude in the context of the FRGE.
Introducing the dimensionless mass $\tilde{m}_k^{2}=k^{-2}{m}_k^{2}$ and expanding for small $\tilde{m}_k^{2}$ we get the standard perturbative and scheme independent result:
\begin{eqnarray*}
\partial_{t}\tilde{m}_{k}^{2} & = & -2\tilde{m}_{\Lambda}^{2}-\frac{1}{(4\pi)^{2}}\frac{\lambda_{\Lambda}}{1+\tilde{m}_{\Lambda}^{2}} =  \left(-2+\frac{\lambda_{\Lambda}}{(4\pi)^{2}}\right)\tilde{m}_{\Lambda}^{2}-\frac{\lambda_{\Lambda}}{(4\pi)^{2}}+...\\
\partial_{t}\lambda_{k} & = & \frac{3}{(4\pi)^{2}}\frac{\lambda_{\Lambda}^{2}}{1+\tilde{m}_{\Lambda}^{2}}= \frac{3\lambda_{\Lambda}^{2}}{(4\pi)^{2}}+...
\end{eqnarray*}
We compute now the finite part of the EAA by integrating the flow of the structure function \eq{flowF}.
Using \eq{Qopt} and (\ref{eve4}) to compute the beta functional of \eq{scalarfrg} we get:
\bea
\partial_{t}\Gamma_{k}[\varphi]
&=&\frac{1}{2}\frac{1}{(4\pi)^{2}}\int d^4x\left\{ \frac{k^6}{k^2+ m^2_{\Lambda}}
-\lambda_{\Lambda}\frac{k^4}{k^2+ m_{\Lambda}^2}\varphi^2\right.\qquad\qquad\nn\\
&&\qquad\qquad\left.+\frac{\lambda_{\Lambda}^2}{4}\frac{k^2}{k^2+m_{\Lambda}^{2}}
\varphi^2\left[1-\sqrt{1-\frac{4k^{2}}{z}}\theta(z-4k^{2})\right]\varphi^2\right\}\,.
\label{4scalarfrg}
\eea
The one-loop effective action $\Gamma_0[\varphi]$ is recovered by integrating 
\eq{4scalarfrg} from $k=\Lambda$ down to $k=0$
\begin{eqnarray}
\Gamma_0[\varphi]&=&\Gamma_\Lambda[\varphi]-\frac{1}{4}\frac{1}{(4\pi)^2}\int d^4x\Bigg\{\left( \frac{1}{2}\Lambda^4-\Lambda^2m_{\Lambda}^2+m^4_{\Lambda}\log\frac{\Lambda^2+m_{\Lambda}^2}{m_{\Lambda}^2}\right)\nonumber\\
&&-\lambda_{\Lambda}\left( \Lambda^2-m_{\Lambda}^2\log\frac{\Lambda^2+m_{\Lambda}^2}{m_{\Lambda}^2}\right)\varphi^2
+\frac{\lambda_{\Lambda}^2}{2}\left(1+\frac{1}{2}\log{\frac{\Lambda^2+m_{\Lambda}^2}{m_{\Lambda}^2}} \right)\varphi^4\nonumber\\
&& -\frac{\lambda_{\Lambda}^2}{2}\varphi^2\sqrt{1+\frac{4m_{\Lambda}^2}{z}}\mbox{ArcTanh}\frac{1}{\sqrt{1+\frac{4m_{\Lambda}^2}{z}}}\varphi^2\Bigg\}\,.
\label{aaa}
\end{eqnarray}
If we try to take the limit $\Lambda\to\infty$ this expression contains quartic, quadratic and logarithmic divergences.
The renormalized action is of the form (\ref{ansatzscalargamma}), with finite ``renormalized'' couplings $E_0$, $m_0$, $\lambda_0$.
The relation between these and the ``bare'' couplings
$E_\Lambda$, $m_\Lambda$, $\lambda_\Lambda$
is contained in renormalization conditions,
which in the present context amount to a choice of
initial condition $\Gamma_\Lambda$.
The finite part of the renormalized
couplings is arbitrary and has to be determined by fitting the
theory to experimental observations.
Here we choose renormalization conditions that simply remove all the
local terms contained in the integral in (\ref{aaa}):
\begin{eqnarray}
E_\Lambda & = & E_0+\frac{1}{4}\frac{1}{(4\pi)^2}\left( \frac{1}{2}\Lambda^4-\Lambda^2m_{\Lambda}^2
+m_{\Lambda}^4\log\frac{\Lambda^2+m_{\Lambda}^2}{m_{\Lambda}^2}\right)
\nonumber\\
m^2_\Lambda & = &m_0^2
-\frac{1}{4}\frac{\lambda_{\Lambda}}{(4\pi)^2} 
\left(\Lambda^2 
-m_{\Lambda}^2\log\frac{\Lambda^2+m_{\Lambda}^2}{m_{\Lambda}^2}\right)
\nonumber\\
\lambda_\Lambda & = &\lambda_0
+\frac{1}{8}\frac{\lambda_{\Lambda}^2}{(4\pi)^2}
\left(1+\log\frac{\Lambda^2
+m_{\Lambda}^2}{m_{\Lambda}^2}\right)\,.
\label{ren_scalar}
\end{eqnarray}
At this point the EA $\Gamma_0$ contains a local part 
of the same form of (\ref{eq:scalar})
except for the replacement of the subcripts $\Lambda$
by subscripts $0$, plus a nonlocal part
that is given by the last line of (\ref{aaa}).
In this part, using the perturbative logic, 
the bare couplings can be also replaced by renormalized ones, 
up to terms of higher order.
It is clear that this step only makes mathematical sense if
$\Lambda$ is bounded (the bound depending on the smallness
of the coupling).
In any case the EA then takes the form
\bea
\Gamma_0[\varphi]  &=& \int d^4x\Bigg\{E_0
+\frac{1}{2}m_0^2\varphi^2
+\frac{1}{4!}\lambda_0\varphi^4
\nonumber\\
&&
+\frac{\lambda_{0}^2}{8(4\pi)^2}\varphi^2\sqrt{1+\frac{4m_{0}^2}{z}}\mbox{ArcTanh}\left(\frac{1}{\sqrt{1+\frac{4m_{0}^2}{z}}}\right)\varphi^2\Bigg\}\,.
\eea
The scattering amplitude for the process $\varphi\varphi\to \varphi\varphi$ is obtained by taking four functional derivatives of the effective action with respect to $\varphi$ after  performing the analytic continuation to Minkowski space. 
Evaluating the expression in Fourier space, we get
\bea\label{ampliscal}
&&A(s,t,u;m_0)=\lambda_0
+\sum_{w=s,t,u}\frac{\lambda_{0}^2}{(4\pi)^2}\sqrt{1
+\frac{4m_{0}^2}{-w}}\mbox{ArcTanh}\frac{1}{\sqrt{1+\frac{4m_{0}^2}{-w}}}
\eea
where $s=(p_1+p_2)^2$, $t=(p_1+p_3)^2$ and $u=(p_1+p_4)^2$
and all momenta are taken to be incoming.

Notice that the expression entering in the r.h.s. of \eq{ampliscal} can be written in terms of the following Feynman integral, which results from the computation of one-loop bubble diagrams: 
\be 
\sqrt{1
+\frac{4m_{0}^2}{z}}\mbox{ArcTanh}\frac{1}{\sqrt{1+\frac{4m_{0}^2}{z}}}=\int_0^1dx\Bigg\{ \frac{1}{2}\log\left[1+\frac{z x (1-x)}{m_0^2} \right]+1\Bigg\}\,.
\ee
For an FRGE computation beyond the one-loop approximation see for example \cite{Codello:2015lba}.

%%%%%%%%%%%%%%%%%%
%%%%%%%%%%%%%%%%%%
\section{Quantum electrodynamics}
%%%%%%%%%%%%%%%%%%
%%%%%%%%%%%%%%%%%%

We now consider Euclidean Quantum Electrodynamics (QED) in $d=4$, which is a perturbatively renormalizable theory characterized by the following  bare action
\be
S[A,\bar{\psi},\psi]=\Gamma_\Lambda[A,\bar{\psi},\psi]=\int d^{4}x\left[\frac{1}{4e_\Lambda^2}F_{\mu\nu}F^{\mu\nu}+\bar{\psi}(\slashed{D}+m_{e})\psi\right]\,,
\label{QED_1}
\ee
where $e_\Lambda$ is the bare electric charge, $m_{e}\equiv m_{e,\Lambda}$ is the bare electron mass, $\slashed{D}=\gamma^{\mu}D_{\mu}$ is the Dirac operator, $D_{\mu}=\partial_{\mu}+iA_{\mu}$ is the covariant derivative and $F_{\mu\nu}$ is the photon field strength tensor 
\be
\Omega_{\mu\nu}=\left[D_{\mu},D_{\nu}\right]=iF_{\mu\nu}=i(\partial_\mu A_\nu-\partial_\nu A_\mu) \,.
\label{QEDfs}
\ee
To quantize the theory we have to introduce a gauge-fixing term which can be taken to be
\be
S_{gf}[A]=\frac{1}{2e^2_\Lambda\alpha}\int d^{4}x\,\left(\partial_{\mu}A^{\mu}\right)^{2}\,,
\label{QEDgfixing}
\ee
where $\alpha$ is the gauge fixing parameter. Notice that the Faddeev-Popov determinant can be safely discarded because on a flat space-time the ghost term decouples.
The one-loop effective action is given by
\bea
\Gamma[A,\bar{\psi},\psi] & = & S[A,\bar{\psi},\psi]+\frac{1}{2}\textrm{Tr}\log\left[-\partial^{2}g^{\mu\nu}+\left(1-\frac{1}{\alpha}\right)\partial^{\mu}\partial^{\nu}-\bar{\psi}\gamma^{\mu}\frac{1}{\slashed{D}+m_{e}}\gamma^{\nu}\psi\right]\nn\\
&&-\textrm{Tr}\log\left(\slashed{D}+m_{e}\right)\,.
\label{QEDea}
\eea
It is useful to rewrite the fermionic trace as
\be
\textrm{Tr}\log\left(\slashed{D}+m_{e}\right)=\frac{1}{2}\log\left[\det\left(\slashed{D}+m_{e}\right)\det\left(-\slashed{D}+m_{e}\right)\right]=\frac{1}{2}\textrm{Tr}\log\left(\Delta+m_{e}^{2}\right)\,,\label{QED_2}
\ee
where
\be 
\Delta=-\slashed{D}^{2}=-D^2-\,\frac{\sigma^{\mu\nu}}{2} F_{\mu\nu}\,\label{QEDfop}
\ee
and $\sigma^{\mu\nu}=\frac{i}{2}\left[\gamma^{\mu},\gamma^{\nu}\right]$. 
We will work in the gauge $\alpha=1$ where the one-loop EAA for QED can be obtained introducing the cutoff kernels directly into \eq{QEDea}
\bea
\Gamma_{k}[A,\bar{\psi},\psi] & = & S[A,\bar{\psi},\psi]+\frac{1}{2}\textrm{Tr}\log\left[-\partial^{2}g^{\mu\nu}-\bar{\psi}\gamma^{\mu}\frac{1}{\slashed{D}+m_{e}}\gamma^{\nu}\psi+R_{k}(-\partial^{2})^{\mu\nu}\right]\nonumber \\
&& -\frac{1}{2}\textrm{Tr}\log\left[\Delta+m_{e}^{2}+R_{k}(\Delta)\right]\,.\label{QEDeaa}
\eea
Notice that in \eq{QEDeaa} we choose the argument of the gauge cutoff function to be the flat-space Laplacian $-\partial^{2}$, while
for the fermion cutoff we take the covariant operator $\Delta$.

The one-loop flow equation is obtained by differentiating \eq{QEDeaa} with respect to the RG parameter $t$.
Here we are interested only in the fermion contribution to the photon effective action $\Gamma_{k}[A]\equiv\Gamma_{k}[A,0,0]$, we have
\begin{eqnarray}
\partial_{t}\Gamma_{k}[A] & = & \frac{1}{2}\textrm{Tr}\frac{\partial_{t}R_{k}(-\partial^{2})^{\mu\nu}}{-\partial^{2}g^{\mu\nu}+R_{k}(-\partial^{2})^{\mu\nu}} -\frac{1}{2}\textrm{Tr}\frac{\partial_{t}R_{k}(\Delta)}{\Delta +m_{e}^{2}+R_{k}(\Delta)}\,.\label{QED1lfe}
\end{eqnarray}
The flow equation for $\Gamma[A]$ is now of the form of \eq{wetterich}.
The first trace in \eq{QED1lfe} does not depend on the photon field
and thus will not generate any $A_{\mu}$ contribution to $\partial_{t}\Gamma_{k}[A]$.
This reflects the fact that QED is an abelian gauge theory with no
photon self-interactions. Thus to one-loop order, all the contributions
to the running of the gauge part of the EAA stem from the fermionic
trace. From now on we will discard the gauge trace.

We calculate the fermion trace in r.h.s. of \eq{QED1lfe} using the non-local heat kernel expansion in \eq{nlhk}.
From \eq{QEDfop} we see that $\Delta$ is the generalized Laplacian operator of \eq{operator} with ${\mathbf U}=-\sigma^{\mu\nu} F_{\mu\nu}/2$.
The function to be traced is
\be
\label{fhqed}
h_{k}(\Delta,m_{e}^{2})=\frac{\partial_{t}R_{k}(\Delta)}{\Delta+m_{e}^{2}+R_{k}(\Delta)}\,\label{QED_5.21}
\ee
We can now specialize \eq{adam} to the QED case:
\bea
\partial_{t}\Gamma_{k}[A] &=&-\frac{1}{2}\frac{1}{(4\pi)^{2}}\int d^{4}x\Big\{ \textrm{tr}\mathbf{1}\int_{0}^{\infty}ds\,s^{-2}\tilde{h}_{k}(s,m_{e}^{2})\nn\\
&&\qquad\qquad\qquad\quad+\,F_{\mu\nu}\left[\int_{0}^{\infty}ds\,\tilde{h}_{k}(s,m_{e}^{2})\, f_{F^{2}}\left(sz\right)\right]F^{\mu\nu}\Big\} \,,
\label{dtqed}
\eea
where $z=-D^2$. The structure function $f_{F^{2}}(x)$ is given by
\footnote{We used $\textrm{tr}\,{\mathbf U}=0$, $\textrm{tr}\,{\mathbf U}^{2}=2F_{\mu\nu}F^{\mu\nu}$ and $\textrm{tr}\,\mathbf{\Omega}_{\mu\nu}\mathbf{\Omega}^{\mu\nu}=-4F_{\mu\nu}F^{\mu\nu}$.}
\be
f_{F^{2}}(x)=2f_{U}(x)-4f_{\Omega}(x)=f(x)+\frac{2}{x}\left[f(x)-1\right]=4\int_{0}^{1}d\xi\,\xi(1-\xi)\, e^{-x\xi(1-\xi)}\,.
\label{sfqed}
\ee
Plugging \eq{sfqed} into \eq{dtqed} and using \eq{shiftq} we get
\be 
\partial_{t}\Gamma_{k}[A] =-\frac{1}{2}\frac{1}{(4\pi)^{2}}\int d^{4}x\left\{ 4\, Q_2[h_k]+4\,F_{\mu\nu}\int_{0}^{1}d\xi\,\xi(1-\xi)\,Q_{0}[h_k^{z\xi(1-\xi)}]F^{\mu\nu}\right\} \,.\label{dtqed2}
\ee
The first constant piece is the renormalization 
of the vacuum energy and we will drop it here.
To proceed we need to specify the form of the ansatz for $\Gamma_{k}[A]$,
to be inserted in the l.h.s. of \eq{dtqed}. We choose:
\be
\Gamma_{k}[A]=\int d^4x\left[\frac{Z_{A,k}}{4}F_{\mu\nu}F^{\mu\nu}+\frac{1}{4}F_{\mu\nu}Z_{A,k}\Pi_{k}\left(-D^{2}\right)F^{\mu\nu}\right]\,, % + O(F^4) \,, 
\label{QEDansatz}
\ee 
where $Z_{A,k}$ is the photon wave function renormalization which is related to the electric charge via the following identification 
\be 
Z_{A,k}=\frac{1}{e_k^2}\,.
\label{zerel}
\ee
The quantity $\Pi_{k}(-D^2)$ is the running photon polarization which is a function of the gauge covariant Laplacian. The $t$-derivative of  \eq{QEDansatz} gives:
\be
\partial_{t}\Gamma_{k}[A]  =  \partial_{t}Z_{A,k}\frac{1}{4}\int d^{4}x\, F_{\mu\nu}F^{\mu\nu} +\frac{1}{4}\int d^{4}x\, F_{\mu\nu}\partial_{t}\left[Z_{A,k}\Pi_{k}\left(-D^{2}\right)\right]F^{\mu\nu}\,.
\label{QEDrantz}
\ee
Comparing \eq{dtqed2} with \eq{QEDrantz} finally gives the flow equation
\be
\partial_{t}Z_{A,k}+\partial_{t}\left[Z_{A,k}\Pi_{k}(z)\right] 
 =  -\frac{1}{2\pi^{2}}\int_{0}^{1}d\xi\,\xi(1-\xi)\,Q_0[h_{k}^{z\xi(1-\xi)}]\,.
 \label{QED_13}
\ee
Since $\Pi_k(0)=0$, the beta function for the wave-function renormalization of the photon field is obtained by evaluating \eq{QED_13} at $z=0$:
\be
\partial_{t}Z_{A,k}=-\frac{1}{2\pi^{2}}\int_{0}^{1}d\xi\,\xi(1-\xi)Q_{0}[h_{k}]
=-\frac{1}{6\pi^{2}}\frac{k^2}{k^2+m_{e}^{2}}\,,
\label{QEDZrun}
\ee
where in the last step we have used \eq{Qopt}. Using the relation in \eq{zerel} we can derive from \eq{QEDZrun} the beta function for the electric charge
\be 
\partial_{t}e^2_{k}=\frac{e_{k}^{4}}{6\pi^{2}}\frac{k^2}{k^2+m_{e}^{2}}\,.\label{QEDbetaf}
\ee
The anomalous dimension of the photon field is given by
\be
\eta_{A,k}=-\frac{\partial_{t}Z_{A,k}}{Z_{A,k}}=\frac{e_{k}^{2}}{6\pi^{2}}\frac{k^{2}}{k^{2}+m_{e}^{2}}\,.\label{QEDeta}
\ee
Notice that in the limit $m_e \ll k$ the fraction in \eq{QEDbetaf} becomes equal to one 
and we recover the standard beta function found in perturbation theory with
a mass independent regularization scheme \cite{pich}.
On the other hand, for $k\ll m_e$ the denominator becomes large and the beta function goes to zero. 
This threshold behavior is the manifestation of the decoupling of the electron at low energy.

If we integrate the beta function for the electric charge in \eq{QEDbetaf} from an UV scale $\Lambda$ down to an IR scale $k$, we find
\begin{equation}
e_{k}^{2}=\frac{e_{\Lambda}^{2}}{1+\frac{e_{\Lambda}^{2}}{12\pi^{2}}\log\frac{1+\Lambda^{2}/m_{e}^{2}}{1+k^{2}/m_{e}^{2}}}\,.\label{QED_eint}
\end{equation}
\Eq{QED_eint} is interesting for several reasons. First,
it shows the screening effect of the vacuum: electron-positron pairs polarize
the vacuum around an electric charge so that the effective electric
charge $e_{k}$, at the scale $k$, is smaller than the electric charge
$e_{\Lambda}$ at the higher scale $\Lambda$. 
Second, for $k\rightarrow0$, it gives the relation between the bare electric charge $e_{\Lambda}$ and the renormalized electric charge $e_{0}$:
\begin{equation}
e_{0}^{2}=\frac{e_{\Lambda}^{2}}{1+\frac{e_{\Lambda}^{2}}{12\pi^{2}}\log\left(1+\frac{\Lambda^{2}}{m_{e}^{2}}\right)}\,.\label{QED_eint0}
\end{equation}
Third, it shows that QED, as defined by the bare action in \eq{QED_1},
is a trivial quantum field theory: if we take the limit $\Lambda\rightarrow\infty$
in \eq{QED_eint0}, at fixed finite $e_{\Lambda}$, we get
a zero renormalized electric charge $e_{0}$.
Conversely, if we solve \eq{QED_eint0} for the bare charge $e_{\Lambda}^{2}$
and we set the renormalized charge $e_{0}^{2}$ to some fixed value,
then the bare coupling will diverge at the finite ``Landau pole''
scale
\be
\Lambda_{L}=m_{e}^{2}\left(e^{12\pi^{2}/e_{0}^{2}}-1\right)\,.
\ee
These are the two faces of QED's triviality. So, even if the theory
is perturbatively renormalizable, it cannot be a fundamental theory
valid at arbitrarily high energy scales. 
To find an explanation for the success of QED, we have to
take the effective field theory point of view.

We now come back to consider the full momentum structure of the r.h.s. of \eq{QED_13}.
Using \eq{zerel} and \eq{QEDZrun} we can read off the running of the photon polarization function
\be
\partial_{t}\Pi_{k}(z)=\frac{e_{k}^2}{6\pi^{2}}
\frac{k^{2}}{k^2+m_{e}^{2}}\left[1+\Pi_{k}(z)\right]
-\frac{e_{k}^{2}}{2\pi^{2}}\int_{0}^{1}d\xi\,\xi(1-\xi)\, Q_0[h_{k}^{z\,\xi(1-\xi)}]\,.
\label{QEDpirun}
\ee
We can find the one-loop renormalized polarization function $\Pi_{0}(x)$ integrating
\eq{QEDpirun} from the UV scale $\Lambda$ down to $k=0$ after having set the coupling $e_k$ to its bare value $e_\Lambda$. Notice that 
the term proportional to $\Pi_{k}(z)$ in the r.h.s. of \eq{QEDpirun} is at least of order
$e^{4}$ and we will discard it in performing the integration since we are interested in reproducing the one-loop result. 
As we did in the section for the real scalar, we use the optimized cutoff of \eq{optimized} to evaluate the $Q$-functional entering in 
\eq{QEDpirun}. Performing the integral we obtain
\be 
\Pi_{\Lambda}(z)-\Pi_{0}(z) =
\frac{e_{\Lambda}^{2}}{2\pi^{2}}\int_{0}^{1}d\xi\,\xi(1-\xi)\,\log\left[1+\xi(1-\xi)\frac{z}{m_{e}^{2}}\right]\,.\label{QEDdipi}
\ee
Notice that \eq{QEDdipi} does not contain divergent pieces,
which are local.

Setting the initial condition $\Pi_\Lambda(z)=0$, we have that the renormalized photon vacuum polarization function is given by
\be
\Pi_{0}(z)=-\frac{e_{\Lambda}^{2}}{2\pi^{2}}\int_{0}^{1}d\xi\,\xi(1-\xi)\log\left[1+\xi(1-\xi)\frac{z}{m_{e}^{2}}\right]\,.\label{QEDpi0}
\ee
Inserting \eq{QEDpi0} in \eq{QEDansatz} and redifining $A_\mu \to A_\mu/Z_A^{1/2}$ we obtain the following one-loop photon effective action
\bea
\Gamma_0[A] & = & \frac{1}{4}\int d^{4}x\, F_{\mu\nu}F^{\mu\nu}
\nonumber \\
&  & 
-\frac{e_\Lambda^{2}}{8\pi^{2}}\int d^{4}x\, F_{\mu\nu}\left(\int_{0}^{1}d\xi\,\xi(1-\xi)\textrm{log}\left[1+\xi(1-\xi)\frac{-D^{2}}{m_{e}^{2}}\right]\right)F^{\mu\nu}
%\nonumber \\
%&  & 
+\,O(F^{4})\,.
\label{QED_ea}
\eea
\Eq{QED_ea} contains all the terms of the QED effective one loop action that are quadratic in the field strength.
%expression for the photon part of
%the EAA in QED to second order in the field strength and in $e_\Lambda^{2}$.
Although the polarization function in \eq{QED_ea} is a function
of $-D^2$, in an abelian theory like QED it boils down to
a function of just the flat Laplacian $-\partial^{2}$ and thus does
not give non-zero contribution to higher vertices of the effective action.

With similar methods one can calculate the local terms in the EAA which are of quartic order in the field strength (and in the derivatives):
\be
\Gamma_k[A]\Big|_{F^4}=a_{k}\int d^{4}x\,\left(F_{\mu\nu}F^{\mu\nu}\right)^{2}+b_{k}\int d^{4}x\, F_{\mu\nu}F^{\nu\alpha}F_{\alpha\beta}F^{\beta\mu}\,,
\label{QED_15}
\ee
where $a_k$ and $b_k$ are the Euler-Heisenberg coefficients with negative quartic mass dimension.
We can compute the fermionic trace in \eq{QED1lfe} using the local heat kernel expansion of \eq{hk}. Contributions of order $F^4$ are given by the coefficient $\mathbf{b}_8(\Delta)$ of the expansion, which, for constant field strength, has the following form \cite{Avramidi:2000bm}:
\be
\mathbf{b}_8(-D^2)= \frac{1}{24}\mathbf{U}^4-\frac{1}{6}\mathbf{U}^2\,\mathbf{\Omega}_{\mu\nu}\mathbf{\Omega}^{\mu\nu} +\left[\frac{1}{288}(\mathbf{\Omega}_{\mu\nu}\mathbf{\Omega}^{\mu\nu})^2
+\frac{1}{360}\mathbf{\Omega}_{\mu\nu}\mathbf{\Omega}^{\nu\alpha}\mathbf{\Omega}_{\alpha\beta}\mathbf{\Omega}^{\beta\mu}\right].
\ee
Using
\be
\textrm{tr}\,\mathbf{U}^4=-\frac{1}{6}F_{\mu\nu}F^{\nu\alpha}F_{\alpha\beta}F^{\beta\mu}+\frac{1}{8}(F_{\mu\nu}F^{\mu\nu})^2\qquad\mbox{and}\qquad
\textrm{tr}\,\mathbf{\Omega}_{\mu\nu}\mathbf{\Omega}^{\nu\alpha}\mathbf{\Omega}_{\alpha\beta}\mathbf{\Omega}^{\beta\mu}
=F_{\mu\nu}F^{\nu\alpha}F_{\alpha\beta}F^{\beta\mu}
\ee
we get
\bea\label{eulerun}
\partial_t a_k &=& \frac{1}{2}\frac{e^4_{\Lambda}}{(4\pi)^{2}}\frac{1}{18}Q_{-2}[h_k] \nn\\
\partial_t b_k &=& -\frac{1}{2}\frac{e^4_{\Lambda}}{(4\pi)^{2}}\frac{7}{45}Q_{-2}[h_k]\,. 
\eea
The Euler-Heisenberg coefficients $a_0$ and $b_0$ entering in the one loop effective action are obtained by integrating \eq{eulerun} from $k=\infty$ down to $k=0$. If we use the mass cutoff shape function for evaluating the $Q$-functionals, we obtain
\bea
a_{0}&=&\frac{1}{36}\frac{e^{4}_{\Lambda}}{(4\pi)^{2}}\int_{0}^{\infty}\frac{dk}{k}\frac{4k^{2}}{\left(k^{2}+m_{e}^{2}\right)^{3}}=\frac{1}{36}\frac{1}{(4\pi)^{2}}\frac{e^{4}_{\Lambda}}{m_{e}^{4}}\nn\\
b_{0}&=&-\frac{7}{90}\frac{e^{4}_{\Lambda}}{(4\pi)^{2}}\int_{0}^{\infty}\frac{dk}{k}\frac{4k^{2}}{\left(k^{2}+m_{e}^{2}\right)^{3}}=-\frac{7}{90}\frac{1}{(4\pi)^{2}}\frac{e^{4}_{\Lambda}}{m_{e}^{4}}\,,
\eea
where we imposed the initial condition $a_{\infty}=b_{\infty}=0$.
These values coincide with the well known result for the Euler-Heisenberg coefficients. 

By plugging these values back into \eq{QED_15} and combining them with the $O(\partial^4)$ term of \eq{QED_ea} we obtain: 
\be
\Gamma_{0}[A]\Big|_{\partial^{4}}=\frac{1}{(4\pi)^{2}}\int d^{4}x
\left[\frac{1}{15}\frac{e^{2}_{\Lambda}}{m_{e}^{2}}F_{\mu\nu}\square F^{\mu\nu}+\frac{1}{36}\frac{e^{4}_{\Lambda}}{m_{e}^{4}}\left(F_{\mu\nu}F^{\mu\nu}\right)^{2}-\frac{7}{90}\frac{e^{4}_{\Lambda}}{m_{e}^{4}}F_{\mu\nu}F^{\nu\alpha}F_{\alpha\beta}F^{\beta\mu}\right]\,.
\label{QEDea4f}
\ee
which is the $p^4$ part of the photon effective action \cite{pich}.
The first term in \eq{QEDea4f} is responsible of the Uehling effect
\cite{uehling}, while the other two describe the low energy scattering of photons mediated by virtual electrons \cite{eh}.
For a non-perturbative use of the FRGE in QED see for example \cite{Gies:2004hy}.

%%%%%%%%%%
%%%%%%%%%%
\section{Yang--Mills}
%%%%%%%%%%
%%%%%%%%%%
The situation for the non-abelian case is quite similar to the abelian one, except for the fact that the gauge bosons are now interacting. We begin by considering the Euclidean Yang--Mills action for the gauge fields $A_\mu^i$ in dimension $d$:
\be
S_{YM}[A]=\frac{1}{4}\int d^dx\, F_{\mu\nu}^i F^{\mu\nu i}\,.
\label{gauge_0.7}
\ee
In \eq{gauge_0.7} the quantity $F^i_{\mu\nu}$ is the gauge field strength tensor defined by
\be 
F^i_{\mu\nu}= \partial_\mu A^i_\nu -\partial_\nu A^i_\mu + i g f^i{}_{jk}A_\mu^j A_\nu^k ,
\ee
where $f^i{}_{jk}$ are the structure constants of the gauge symmetry group and $g$ is the coupling constant. The EAA is constructed using the background field method \cite{Reuter:1993kw,Gies:2002af}. The gauge field is split as follows
\be 
A_\mu=\bar A_\mu+a_\mu\,,
\ee
where $a_\mu$ parametrizes the gauge fluctuations around the background field $\bar A_\mu$. In the following we will remove the bar and we will denote the background field simply by $A_\mu$.
In order to properly quantize the theory we choose as a gauge fixing condition $\chi^i=D_\mu a^{\mu i}$, where $D$ is the covariant derivative constructed with the background connection acting on fields in the adjoint representation. The gauge fixing action then reads
\be
S_{gf}[a;A]=\frac{1}{2\alpha}\int d^d x D_\mu a^{\mu i} D_\nu a^{\nu i}\,,
\label{gf}
\ee
where $\alpha$ is the gauge fixing parameter.
The background ghost action reads
\be
S_{gh}[a,\bar{c},c;A] = \int d^d x \hat D_\mu\bar{c}^i\,D^\mu c^i\,,
\label{gh}
\ee
where $\bar{c}$ and $c$ are the ghost fields
and $\hat D$ is the covariant derivative constructed with the full field. 
The total action is then obtained by summing the three contributions
\be\label{ymtotact}
S[a,\bar{c},c;A]=S_{YM}[A+a]+S_{gf}[a;A]+S_{gh}[a,\bar{c},c;A] \,.
\ee 
The background effective action $\Gamma_k[a,\bar{c},c;A]$ which is constructed using the background field method is a functional of the background field ($A_\mu$) and of the classical fields conjugated to the currents coupled to the quantum fluctuations, which we denote again by ($a_\mu,\bar{c},c$). The background EA is invariant under the simultaneous gauge transformation of both.
One can define the gauge invariant EAA by setting the classical fields to zero $\Gamma_k[A]\equiv \Gamma_k[0,0,0;A]$. In the following we will study the RG flow of $\Gamma_k[A]$. 
(Note that in this case $\hat D$ can be replaced by $D$
in the ghost action.)

The exact RG equation for $\Gamma_k[A]$ can be found in \cite{Codello:2013wxa}.
Using \eq{ymtotact} we obtain
\be
\partial_{t}\Gamma_{k}[A] = \frac{1}{2}\textrm{Tr}\frac{\partial_{t}R_{k}(D_T)}{D_T+R_{k}(D_T)} -\frac{1}{2}\textrm{Tr}\frac{\partial_{t}R_{k}(-D^{2})}{-D^{2}+R_{k}(-D^{2})}\,,
\label{flow_YM}
\ee
where the gauge-covariant Laplacian is  $(D_T)^{ij\,\mu\nu}\equiv(-D^2)^{ij}\eta^{\mu\nu}+\mathbf{U}^{ij\,\mu\nu}$, with  $\mathbf{U}^{ij\,\mu\nu}=2f^{ij}{}_lF^{l\,\mu\nu}$. The commutator of covariant derivatives is $[D_\mu,D_\nu]^{ij}=\mathbf{\Omega}_{\mu\nu}^{ij}=-f^{ij}{}_l F^l_{\mu\nu}$.

We can now use the non--local heat kernel expansion of \eq{adam} to compute  the r.h.s. of \eq{flow_YM}. For the gauge group $SU(N)$, we get:
\be
\partial_{t}\Gamma_{k}[A]\Big|_{F^{2}}=\frac{N}{(4\pi)^{d/2}}\int d^{d}x\,F_{\mu\nu}^{i}\left[\int_{0}^{\infty}ds\,\tilde{h}_{k}(s,0)\, s^{2-d/2}\, f_{F^{2}}\left(-sD^{2}\right)\right]F^{i\mu\nu}\,,
\label{gauge_0.16}
\ee
where the structure function $f_{F^{2}}(x)$ is given by:
\be
f_{F^{2}}(x)=\frac{1}{2}[4f_U(x)-df_\Omega(x)]+f_\Omega(x)=f(x)+\frac{d-2}{4x}\left[f(x)-1\right]\,.\label{gauge_0.17}
\ee
We need now to make an ansatz for the l.h.s. of the flow equation.
Retaining terms up to second order in the field strength,
but with arbitrary momentum dependence, the EAA has the form:
\be
\Gamma_{k}[A]=\frac{Z_{A,k}}{4}\int d^{d}x\, F_{\mu\nu}^{a}F^{a\mu\nu}+\frac{Z_{A,k}}{4}\int d^{d}x\, F_{\mu\nu}^{a}\Pi_{k}\left(-D^{2}\right)^{ab}F^{b\mu\nu}+O\left(F^{3}\right)\,,\label{gauge_0.15}
\ee
where $Z_{A,k}=1/g_k^{2}$ and $\Pi_{k}(z)$ is the running vacuum polarization function.
%
%If we expand \eq{gauge_0.15} in powers of the fluctuation field we find to lowest order
%
%\be
%\Gamma[Z_{A,k}^{1/2}A+a]=Z_{A,k}\Gamma[A]+O(a)\,,
%5\label{gauge_0.151}
%\ee
%
%since the background covariant Laplacian does not renormalize and .
Notice that the background wave--function renormalization constant, and so the gauge coupling,
enters the flow of $\Pi_{k}(x)$ only as an overall factor.

Comparing the expression of \eq{gauge_0.15} with \eq{gauge_0.16} and using \eq{adam2}, we get
the flow equation for the running vacuum polarization function:
\bea
\partial_{t}Z_{A,k}+\partial_{t}\left[Z_{A,k}\Pi_{k}(z)\right] &=& 
\frac{N}{(4\pi)^{d/2}}\Bigg\{ 4\int_{0}^{1}d\xi\, Q_{\frac{d}{2}-2}[h_k^{z\xi(1-\xi)}]\nn\\
&&+\frac{d-2}{z}\left(\int_{0}^{1}d\xi\, Q_{\frac{d}{2}-1}[h_k^{z\xi(1-\xi)}]
-Q_{\frac{d}{2}-1}[h_{k}]\right)\Bigg\} \,,
\label{gauge_0.19}
\eea
where $z$ stands for the covariant Laplacian $-D^2$.
Evaluating the $Q$--functionals for the optimized cutoff (\ref{optimized}) 
in dimension $d=4$ gives:
\be
\partial_{t}Z_{A,k}+\partial_{t}\left[Z_{A,k}\Pi_{k}(z)\right]=\frac{N}{(4\pi)^{2}}\left[\frac{22}{3}-\left(\frac{22}{3}+\frac{8k^{2}}{3z}\right)\sqrt{1-\frac{4k^{2}}{z}}\theta(z-4k^{2})\right]\,.\label{gauge_0.20}
\ee
Since $\Pi_k(0)=0$ we can rewrite the above equation as
\be
\eta_{A,k}=-\frac{\partial_t Z_{A,k}}{Z_{A,k}}=-\frac{22}{3}\frac{N}{(4\pi)^{2}}g^2_k
\label{eta_YM}
\ee
and
\be
\partial_{t}\Pi_{k}(z)=\eta_{A,k}\Pi_{k}(z)+\frac{g_{k}^{2}N}{(4\pi)^{2}}g_{F^2}\left(\tilde z\right)\,.
\label{gauge_0.2001}
\ee
The function $g_{F^2}(\tilde z)$ is:
\be
g_{F^2}(\tilde z)=-\left(\frac{22}{3}+\frac{8}{3\tilde z}\right)\sqrt{1-\frac{4}{\tilde z}}\,\theta(\tilde z-4)\,.
\label{gauge_0.201}
\ee
Notice that the $k$ dependence in \eq{gauge_0.2001} enters only via the combination $\tilde z=z/k^{2}$ and
%
%As usual $\eta_{A,k}=-\partial_t \log Z_{A,k}$ and 
we used the relation $g_{k}^{2}=Z_{A,k}^{-1}$.

From the anomalous dimension in \eq{eta_YM} we find immediately the beta function for the gauge coupling:
\be
\partial_{t}g_{k}=\frac{1}{2}\eta_{A,k}g^2_k=-\frac{N}{(4\pi)^{2}}\frac{11N}{3}g_{k}^{3}\,,
\label{gauge_0.13}
\ee
which is the standard one--loop result.
Integrating the one-loop beta function for the gauge coupling
from the UV scale $\Lambda$ to the IR scale $k$ we find
\be
g_{\Lambda}^{2}=\frac{g_{k}^{2}}{1+\frac{g_{k}^{2}}{(4\pi)^{2}}\frac{22N}{3}\log\frac{\Lambda}{k}}\,.
\label{gauge_0.143}
\ee
A mass scale $M$ can be defined by the relation
\be
1=\frac{g_{k}^{2}}{(4\pi)^{2}}\frac{22N}{3}\log\frac{k}{M}\,,
\label{gauge_0.144}
\ee
if we insert \eq{gauge_0.144} in \eq{gauge_0.143} we can write
\be
\alpha_{\Lambda}=\frac{2\pi}{\frac{11}{3}N\,\log\frac{\Lambda}{M}}\,,
\label{gauge_0.146}
\ee
where we defined $\alpha_{\Lambda}=\frac{g_{\Lambda}^{2}}{4\pi}$.
This is the standard result found in perturbation theory.

We now go back to the running of the vacuum polarization in \eq{gauge_0.2001}. The term $\eta_{A,k}\Pi_{k}(z)$ is at least of order $g_{k}^{4}$ and we will discard it here since we are interested in reproducing the one-loop result.
Moreover, we set the running coupling to its bare value $g_{\Lambda}$.
We can now integrate the flow of $\Pi_{k}(z)$ in \eq{gauge_0.2001} from an UV scale $\Lambda$ down to an IR scale $k$. We get:
\be
\Pi_{\Lambda}(z)-\Pi_{k}(z)=\frac{g_{\Lambda}^{2}N}{2(4\pi)^{2}}\int_{z/\Lambda^{2}}^{z/k^{2}}\frac{du }{ u}\, g_{F^2}(u)\ .\label{gauge_0.202}
\ee
The integral in \eq{gauge_0.202} is finite in the limit $\Lambda\rightarrow\infty$ and no renormalization is needed.
The function $g_{F^2}$ has no constant term so for every $z$ and $k$ big enough the flow of $\Pi_k(z)$ is zero and no divergences can develop. 
In this limit, the vacuum polarization function goes to its boundary value, i.e. $\Pi_{\Lambda}(z)=0$.
Using the general integrals of eqs. (\ref{HK_Q_9})--(\ref{HK_Q_11}), the vacuum polarization function at the scale $k$ is finally found to be:
\be
\Pi_{k}(z) = -\frac{g_{\Lambda}^{2}N}{(4\pi)^{2}}\left\{
-\frac{22}{3}\left[\frac{1}{2}\log\frac{z}{k^{2}}+\log\frac{1+\sqrt{1-\frac{4k^{2}}{z}}}{2}\right]
 +\left(\frac{64}{9}+\frac{8k^{2}}{9z}\right)\sqrt{1-\frac{4k^{2}}{z}}\right\}
\label{gauge_0.203}
\ee
for $z/k^2\geq 4$ and $\Pi_{k}(z) =0$ for $z/k^2 < 4$.
From \eq{gauge_0.203} we see that we cannot send $k\rightarrow0$, since the first logarithm diverges in this limit. 
For $k^{2}\ll z$, \eq{gauge_0.203} gives the following contribution to the gauge invariant EAA:
\be
\frac{g_{\Lambda}^{2}N}{64\pi^{2}}\int d^{4}x\, F_{\mu\nu}^{i}\left[\frac{11}{3}\log\frac{(-D^{2})^{ab}}{k^{2}}-\frac{64}{9}\delta^{ab}\right]F^{i\mu\nu}\,.\label{gauge_0.204}
\ee
We can interpret the obstruction to taking the limit $k\rightarrow0$ in \eq{gauge_0.203}
as a signal of the breakdown of the approximation used in its derivation,
where we considered the flow of $\Pi_{k}(z)$ as driven only by the
operator $\frac{1}{4}\int F^{2}$. In order to be able to continue
the flow of the EAA in the deep IR, we need the full non-perturbative
power of the exact RG flow equation that becomes available
if we insert the complete ansatz (\ref{gauge_0.15}) in the r.h.s. side of it \cite{Ellwanger_Hirsch_Weber_1998}.

%%%%%%%%%%%%%%%%%%%%%%%%%%%%%%%%%%%%%%%%%%%%%%%%%%
\section{The chiral model} 
%%%%%%%%%%%%%%%%%%%%%%%%%%%%%%%%%%%%%%%%%%%%%%%%%%

In the previous sections we have considered perturbatively renormalizable theories.
In the remaining two we shall consider non-renormalizable ones.
The standard way of treating non-renormalizable theories is the method of effective field theories \cite{pich}.
We shall see here how to recover some well--known results
of the EFT approach using the FRGE.
Previous application of the FRGE to the nonlinear sigma models
have been discussed in \cite{codello,zanusso,fptz,fptv,bfptv,fwz}
The dynamics of Goldstone bosons is described by the nonlinear sigma model,
a theory of scalar fields with values in a homogeneous space.
In particular in QCD with $N$ massless quark flavors the Goldstone bosons of spontaneously broken
$SU_L(N)\times SU_R(N)$ symmetry correspond to the meson multiplet.
These theories are known as the chiral models. 
They have derivative couplings and their perturbative expansion 
is ill-defined in the UV.
A related and phenomenologically even more pressing issue is the high energy behavior of the
tree level scattering amplitude, which grows like $s/F_\pi^2$, where $s$ is the c.m. energy squared
and $F_\pi$ (the ``pion decay constant'') is the inverse of the perturbative coupling.
This leads to violation of unitarity at energies of the order $\sim 4\pi F_\pi$,
which is usually taken as the first sign of the breakdown of the theory.

The chiral NLSM that we consider here is a theory of three scalar fields $\pi^\alpha(x)$,
called the ``pions'', parameterizing (in a neighborhood of the identity) the group $SU(2)$.
Geometrically, they can be regarded as normal coordinates on the group.
We call $U$ the fundamental representation of the group element
corresponding to the field $\pi^\alpha$:
$U=\exp(f\pim)$, $\pim=\pi^\alpha T_\alpha$, $T_\alpha=\frac{i}{2}\sigma_\alpha$, $T_\alpha^\dagger=-T_\alpha$, $\tr(T_\alpha T_\beta)=-\frac{1}{2}\delta_{\alpha\beta}$, $\alpha=1,2,3$.
The coupling $f$ is related to the pion decay constant as $F_\pi=2/f$.
The standard $SU(2)_L\times SU(2)_R$-invariant action for the chiral model is\footnote{Since $U^{-1}=U^\dagger$ we have that $\tr U^{-1} \partial_{\mu} U \, U^{-1} \partial^{\mu} U = -\tr \partial_{\mu} U\,\partial^{\mu} U^\dagger $.}
\be
S[\pi]=-\frac{1}{f^2_{\Lambda}}\int d^4x\,\tr U^{-1}\partial_{\mu} U\,U^{-1}\partial^{\mu} U \, .
\ee
This is the term with the lowest number of derivatives.
Terms with more derivatives will be discussed later.
Introducing the above formulae and keeping terms up to six powers of $\pi$'s we get
\bea
S[\pi]&=&\frac{1}{2}\int d^4x\,
\Bigg\{(\partial_\mu\pi^\alpha)^2-\frac{1}{12}f^2_{\Lambda}\left[\pi^\alpha\pi_\alpha(\partial_\mu\pi^\beta)^2-(\pi^\alpha\partial_\mu \pi_\alpha)^2\right]\nn\\
&&\qquad\qquad+\frac{1}{360}f^4_{\Lambda} \left[(\pi^\alpha\pi_\alpha)^2(\partial_\mu\pi^\beta)^2-\pi^\alpha\pi_\alpha(\pi^\beta\partial_\mu \pi_\beta)^2\right]+O(\pi^8)\Bigg\}\,.
\label{actionNLSM}
\eea
If we define the dimensionless fields $\varphi^\alpha=f\pi^\alpha$ and the metric
\be
\label{metric}
h_{\alpha\beta}(f_{\Lambda}\pi)=\delta_{\alpha\beta}-\frac{1}{12}f^2(\pi^\sigma \pi_\sigma\delta_{\alpha\beta}-\pi_\alpha\pi_\beta)
+\frac{1}{360}f^4\,[(\pi^\sigma\pi_\sigma)^2 \delta_{\alpha\beta}-\pi^\sigma\pi_\sigma\pi_\alpha\pi_\beta]+O(\pi^8)\,,
\ee
we can rewrite \eq{actionNLSM} as a non--linear sigma model:
\be
S[\varphi]=\frac{1}{2f^2}\int d^4x\,
h_{\alpha\beta}(\varphi)\partial_\mu\varphi^\alpha\partial^\mu\varphi^\beta\,.
\label{nlsm}
\ee
Note that the pion fields $\pi^\alpha$ are canonically normalized
and the metric is dimensionless, whereas the fields $\varphi^\alpha$ are dimensionless.

Following \cite{honerkamp} we use the background field method and expand a field $\varphi$ around a
background $\bar\varphi$ using the exponential map:
$\varphi(x)=\exp_{\bar\varphi(x)}\xi(x)$ where the quantum field $\xi$ 
is geometrically a vector field along $\bar\varphi$.
The EAA will be, in general, a function of the background field and the Legendre transform of sources coupled linearly to $\xi$,
which we will denote by the same symbol hoping that this will cause no confusion.
We also omit the bar over the background field so that we can write the EAA as $\Gamma_k[\xi;\varphi]$.
For our purposes it will be sufficient to compute this EAA at $\xi=0$: $\Gamma_k[\varphi]\equiv\Gamma_k[0;\varphi]$.
The RG flow for $\Gamma_k[\varphi]$ is driven by the Hessian 
$\Gamma_k^{(2)}[\varphi]$.
In the one-loop approximation that we shall use, this is equal to
\be
S^{(2)}[\varphi]_{\alpha \beta} = \frac{1}{f^2} 
\left(-\square \,h_{\alpha \beta}+U_{\alpha \beta} \right)\ ,
\label{BFaction}
\ee
where $\square \equiv \nabla_\mu\nabla^\mu$, $\nabla_\mu$ is the covariant derivative with respect to the Riemannian connection of $h_{\alpha\beta}$, $U_{\alpha\beta}=- R_{\epsilon\alpha\eta\beta}
\partial_\mu\varphi^\epsilon\partial^\mu\varphi^\eta$.
We have expressed the second variation in terms of the dimensionless background fields $\varphi^\alpha$, which produces the overall factor $1/f^2$. In the one-loop approximation the running of couplings
in the r.h.s. of the FRGE is neglected and $f$ 
has to be kept fixed along the flow.
Since geometrically $SU(2)$ is a three--sphere with radius two \cite{zanusso}, the Riemann tensor can be written in the form $R_{\alpha\beta\gamma\delta}=\frac{1}{2}(h_{\alpha\gamma}h_{\beta\delta}-h_{\alpha\delta}h_{\beta\gamma})$.
The appearance of the covariant Laplacian suggests the choice of $\Delta=-\square+U$ as argument of the cutoff kernel function.
In this way the cutoff combines with the quadratic action to produce the
function $h_k(\Delta,\omega)$ given in \eq{function}, with $\omega=0$.

Evaluation of the trace follows the steps outlined in section II and one arrives at
the beta functional:
\bea
\partial_t \Gamma_k[\pi]&=&
\frac{1}{2}\frac{1}{(4\pi )^2}\int d^4x \Biggl\{
k^2f^2\,\partial_\mu\pi_\alpha \partial^\mu\pi^\alpha
-\frac{1}{12}k^2f^4\,
\left(
\partial_\mu\pi_\alpha \partial^\mu\pi^\alpha \pi_\beta \pi^\beta
-
\partial_\mu\pi^\alpha \partial^\mu\pi^\beta \pi_\alpha \pi_\beta \right)
\nn\\
&&+\frac{1}{16}f^4\left[\partial_\mu\pi^\alpha \partial^\mu\pi_\alpha
\,g_U (-\square/k^2)
\partial_\nu\pi^\beta \partial^\nu\pi_\beta
+\partial_\mu\pi^\alpha\partial^\mu\pi _\beta
\,g_U (-\square/k^2)
\partial_\nu\pi^\beta \partial^\nu\pi_\alpha\right]
\nn\\
&&+\frac{1}{8}f^4\left[\partial^\mu\pi^\beta \partial_\nu\pi_\alpha 
\,g_{\Omega}  (-\square/k^2)
\partial_\mu\pi^\alpha \partial^\nu\pi_\beta
-\,\partial^\mu\pi_\alpha \partial_\nu\pi^\beta 
\,g_{\Omega} (-\square/k^2)
\partial_\mu\pi^\alpha \partial^\nu\pi_\beta\right]\nn\\
&&+O(\pi^6)\Biggr\}\ ,
\label{rhsnlsm}
\eea
where we have used $\Omega_{\mu\nu\,\alpha\beta}=\partial_\mu\varphi^\epsilon\partial_\nu\varphi^\eta R_{\alpha\beta\epsilon\eta}$ and eqs.~(\ref{eve4})--(\ref{eve5}).
The form factors $g_U(-\square/k^2)$ and $g_\Omega(-\square/k^2)$
correspond to $g_U(-\square,0,k)$ and $g_\Omega(-\square,0,k)$
in the notation of Section II.
It is important to stress that in the derivation of this result no regularization was needed: 
the integrals we had to perform were IR and UV finite.
This is a general property of the beta functional $\partial_t\Gamma_k$.

Another important fact to note is that the first two terms
appear in the same ratio as in the original action of \eq{actionNLSM}.
This is just a consequence of the fact that the cutoff preserves the 
$SU_L(2)\times SU_R(2)$ invariance of the theory. 
As a result, in the RG flow the metric $h_{\alpha\beta}(\varphi)$ gets only rescaled by an overall
factor when expressed in terms of the dimensionless field $\varphi$.

We see that in addition to terms of the same type of the 
original action,
quantum fluctuations generate new terms with four derivatives 
of the fields.
To this order one can write an ansatz for the EAA that contains generic
four-derivative terms:
\bea
\label{eaa}
\Gamma_k[\pi]&=&
\int d^4x\,\Biggl\{
\frac{1}{2}
\partial_\mu\pi_\alpha \partial^\mu\pi^\alpha
-\frac{1}{24}f_k^2\left(
\partial_\mu\pi_\alpha \partial^\mu\pi^\alpha \pi_\beta \pi^\beta
-
\partial_\mu\pi^\alpha \partial^\mu\pi^\beta \pi_\alpha \pi_\beta \right)
\nn\\
&&+\,\partial_\mu\pi_\alpha \partial^\mu\pi^\alpha 
f_k^4\gamma_{U,k}(\square) \partial_\nu\pi_\beta \partial^\nu\pi^\beta
+\,\partial_\mu\pi^\alpha\partial^\mu\pi^\beta 
f_k^4\gamma_{U,k}(\square) \partial_\nu\pi_\alpha \partial^\nu\pi_\beta
\nn\\
&&
-\,\partial^\mu\pi^\alpha \partial^\nu\pi^\beta 
f_k^4\gamma_{\Omega,k}(\square) \partial_\mu\pi_\alpha \partial_\nu\pi_\beta
+\,\partial^\mu\pi^\beta \partial^\nu\pi^\alpha 
f_k^4\gamma_{\Omega,k}(\square) \partial_\mu\pi_\alpha \partial_\nu\pi_\beta
\nn\\
&&+O(\pi^6)\Biggr\} \,.
\eea
Here the coupling $f_k$ and the dimensionless form factors $\gamma_{U,k}$ and $\gamma_{\Omega,k}$ are functions of the scale $k$.
The local parts of the form factors $\gamma_{U,k}(0)$ and $\gamma_{\Omega,k}(0)$ can be combined as:
\bea
\label{eaanlsmcff}
\Gamma_k[\pi]&=&
\int d^4x\,\Biggl\{
\frac{1}{2}
\partial_\mu\pi_\alpha \partial^\mu\pi^\alpha
-\frac{1}{24}f_k^2\left(
\partial_\mu\pi_\alpha \partial^\mu\pi^\alpha \pi_\beta \pi^\beta
-
\partial_\mu\pi^\alpha \partial^\mu\pi^\beta \pi_\alpha \pi_\beta \right)
\nn\\
&&+\,\frac{\ell_{1,k}}{2}f_k^4\,\partial_\mu\pi_\alpha \partial_\nu\pi^\alpha 
 \partial^\mu\pi_\beta \partial^\nu\pi^\beta
+\frac{\ell_{2,k}}{2}f_k^4\,\partial_\mu\pi_\alpha \partial^\mu\pi^\alpha 
 \partial_\nu\pi_\beta \partial^\nu\pi^\beta +O(\pi^6)\Biggr\} \,,
\eea
where
\be \label{ellllls}
\frac{\ell_{1,k}}{2}=\gamma_{U,k}(0)+\gamma_{\Omega,k}(0)\qquad\mbox{and}\qquad \frac{\ell_{2,k}}{2}=\gamma_{U,k}(0)-\gamma_{\Omega,k}(0)\,.
\ee
Up to this point we have interpreted an infinitesimal 
RG transformation as the result of an integration over 
an infinitesimal shell of momenta $a k<p<k$, with $a=1-\epsilon$ and $\epsilon>0$.
In the literature on the Wilsonian RG, however, this is complemented
by two additional transformations: a rescaling of all momenta
by a factor $a$ and a rescaling of all fields so that they remain
canonically normalized.
The rescaling of the momenta is effectively taken into account by
the rescaling of the couplings $\tilde f_k=k f_k$, etc.
We will now explicitly implement the rescaling of the fields:
\be
\label{etapi}
\delta\pi^\alpha=
-\frac{1}{2}\frac{1}{(4\pi)^2}\tilde f^2\,\pi^\alpha\delta t\ ,\quad
\ee
Defining $\hat\partial_t$ by
$$
\hat\partial_t=\partial_t
+\frac{\delta\pi^a}{\delta t}\frac{\delta}{\delta\pi^a}
$$ 
we obtain
\bea
\hat\partial_t \Gamma_k[\pi]&=&
\frac{1}{(4\pi )^2}\int d^4x \Biggl\{
\frac{1}{24}k^2f^4\,
\left(
\partial_\mu\pi_\alpha \partial^\mu\pi^\alpha \pi_\beta \pi^\beta
-
\partial_\mu\pi^\alpha \partial^\mu\pi^\beta \pi_\alpha \pi_\beta \right)
\nn\\
&&+f^4\Bigg[\partial_\mu\pi^\alpha \partial^\mu\pi_\alpha
\left(\frac{1}{32}g_U (-\square/k^2)-2\tilde f^2\gamma_{U,k}(\square)\right)
\partial_\nu\pi^\beta \partial^\nu\pi_\beta
\nn\\
&&
\qquad
+\partial_\mu\pi^\alpha\partial^\mu\pi _\beta
\left(\frac{1}{32}g_U (-\square/k^2)-2\tilde f^2\gamma_{U,k}(\square)\right)
\partial_\nu\pi^\beta \partial^\nu\pi_\alpha
\nn\\
&&
\qquad
+\partial^\mu\pi^\beta \partial_\nu\pi_\alpha 
\left(\frac{1}{16}g_{\Omega}(-\square/k^2)-2\tilde f^2\gamma_{\Omega,k}(\square)\right)
\partial_\mu\pi^\alpha \partial^\nu\pi_\beta
\nn\\
&&
\qquad-\partial^\mu\pi_\alpha \partial_\nu\pi^\beta 
\left(\frac{1}{16}g_{\Omega}(-\square/k^2)-2\tilde f^2\gamma_{\Omega,k}(\square)\right)
\partial_\mu\pi^\alpha \partial^\nu\pi_\beta\Bigg]
+O(\pi^6)\Biggr\}\,.\nn\\
\label{rhsnlsmhat}
\eea
Comparing the $t$-derivative of \eq{eaa} with \eq{rhsnlsmhat} we get that, by construction, the kinetic term does not change along the RG flow while
for the second term we find instead
\be
\label{betaf2}
\hat\partial_t f^2_k=-\frac{k^2 f^4}{16\pi^2}\ .
\ee
For the nonlocal form factors, one obtains:
\bea
\hat\partial_t\gamma_{U,k}(\square)&=&
\frac{1}{16\pi^2}\frac{1}{32}\,g_U (-\square/k^2)\,,\nn\\
\hat\partial_t \gamma_{\Omega,k}(\square)&=&
\frac{1}{16\pi^2}\frac{1}{16}\,g_{\Omega}(-\square/k^2)\,.
\label{betagamma2}
\eea
In order to recover the standard perturbative result for the effective action we integrate \eq{rhsnlsmhat} from some initial UV scale $k=\Lambda$,
which we can view as the ``UV cutoff'', down to $k=0$, keeping the Goldstone coupling $f$ fixed at the value it has in the bare action ($f_\Lambda$) and 
neglecting corrections of order ${\cal O}(f^4\tilde f^2)$ to the flow of the form factors. In fact, from the integration of \eq{betaf2} we will see that 
the coupling $f^2$ changes by a factor $1+\frac{\tilde f_\Lambda^2}{32\pi^2}$, which is a number close to one in the domain of validity
of the effective field theory (even if $\Lambda\approx f^{-1}$, 
$f$ changes only by a few percent).
Thus we have
\bea 
\Gamma_\Lambda-\Gamma_0&=&\int_0^\Lambda \frac{dk}{k} \hat\partial_t \Gamma_k[\pi]\nn\\
&=&\int_0^\Lambda \frac{dk}{k}\int d^4x\, \Biggl\{-\frac{1}{24}\hat\partial_t f^2_k\left(
\partial_\mu\pi_\alpha \partial^\mu\pi^\alpha \pi_\beta \pi^\beta
-
\partial_\mu\pi^\alpha \partial^\mu\pi^\beta \pi_\alpha \pi_\beta \right)\\
&&\quad +\, f^4\Bigg[\partial_\mu\pi^\alpha \partial^\mu\pi_\alpha
\hat\partial_t\gamma_{U,k}(\square)
\partial_\nu\pi^\beta \partial^\nu\pi_\beta+\partial_\mu\pi^\alpha\partial^\mu\pi_\beta
\hat\partial_t\gamma_{U,k}(\square)\partial_\nu\pi^\beta \partial^\nu\pi_\alpha
\nn\\
&&\qquad \quad +\partial^\mu\pi^\beta \partial_\nu\pi_\alpha \hat\partial_t\gamma_{\Omega,k}(\square)
\partial_\mu\pi^\alpha \partial^\nu\pi_\beta
-\partial^\mu\pi_\alpha \partial_\nu\pi^\beta \hat\partial_t\gamma_{\Omega,k}(\square)
\partial_\mu\pi^\alpha \partial^\nu\pi_\beta\Bigg]\,.
%\nn\\
%&&
%\qquad \quad 
+O(\pi^6)\Biggr\}
\nn
\eea
The effective action is then obtained by integrating  \eq{betaf2} and \eq{betagamma2}. We get
\be
\label{int1}
f^2_\Lambda-f^2_0=\int_0^\Lambda \frac{dk}{k} \hat\partial_t f_k^2
=-\frac{1}{2}\frac{f_\Lambda^4}{16\pi^2}\Lambda^2
\ee
and
\be
\gamma_{U,\Lambda}(\square)-\gamma_{U,0}(\square)=
\int_0^\Lambda \frac{dk}{k} \hat\partial_t\gamma_{U,k}(\square)
=\frac{1}{32}\frac{1}{16\pi^2}\left(1-\frac{1}{2}\log\frac{-\square}{\Lambda^2}\right)\,,
\label{int2}
\ee
\be
\gamma_{\Omega,\Lambda}(\square)-\gamma_{\Omega,0}(\square)=
\int_0^\Lambda \frac{dk}{k} \hat\partial_t\gamma_{k,\Omega}(\square)
=\frac{1}{16}\frac{1}{16\pi^2}\left( \frac{2}{9}-\frac{1}{12}\log\frac{-\square}{\Lambda^2}\right)\,.
\label{int3}
\ee

In this effective field theory approach the need to renormalize 
does not arise
so much from having to eliminate divergences, since $\Lambda$ corresponds to
some finite physical scale and all integrals are finite anyway.
Instead, it is dictated by the desire to eliminate all dependence on high energy couplings,
which are unknown or only poorly known, and to express everything in terms of quantities
that are measurable in the low energy theory.
\footnote{
Ultimately, this produces the same effect, for when the high energy parameters have been
eliminated one may as well send $\Lambda$ to infinity.}
So, in order to eliminate the quadratic dependence on $\Lambda$, we define the renormalized coupling to be:
\be
f_0^2=f_\Lambda^2+\frac{f_\Lambda^4}{32\pi^2}\Lambda^2\,.
\ee
To get rid of the logarithmic dependence on $\Lambda$ we define the renormalized values of $\ell_1$ and $\ell_2$, introduced in \eq{ellllls}, to be:
\be
\ell_{1,0}=\ell_{1,\Lambda}+\frac{4}{3}\frac{f_\Lambda^4}{512\pi^2}\log\frac{\mu^2}{\Lambda^2}\ ,
\qquad
%\ee
%
%\be
\ell_{2,0}=\ell_{2,\Lambda}+\frac{2}{3}\frac{f_\Lambda^4}{512\pi^2}\log\frac{\mu^2}{\Lambda^2}\,.
\ee
Here the couplings with subscripts $\Lambda$ can be interpreted as the ``bare'' couplings
and the ones with subscripts $0$ can be interpreted as the ``renormalized'' ones.
Notice that, once subtracted, the quadratic dependence completely disappears, but due to the need to
compensate the dimension of the Laplacian in the logarithms, one has to define the
renormalized $\ell_1$ and $\ell_2$ at some (infrared) scale $\mu$
and therefore there unavoidably remains a residual scale dependence.

After this renormalization procedure, the effective action can be written, in perturbation theory, as a function of the renormalized coupling:
\bea \label{renactionnlsm}
\Gamma_0[\pi]&=&\int d^4x\,\Biggl\{
\frac{1}{2}
\partial_\mu\pi_\alpha \partial^\mu\pi^\alpha
-\frac{1}{24}f^2\left(
\partial_\mu\pi_\alpha \partial^\mu\pi^\alpha \pi_\beta \pi^\beta
-
\partial_\mu\pi^\alpha \partial^\mu\pi^\beta \pi_\alpha \pi_\beta \right)
\nn\\
&&\qquad\quad+\,\frac{\ell_{1}}{2}f^4\,\partial_\mu\pi_\alpha \partial_\nu\pi^\alpha 
 \partial^\mu\pi_\beta \partial^\nu\pi^\beta
+\frac{\ell_{2}}{2}f^4\,\partial_\mu\pi_\alpha \partial^\mu\pi^\alpha 
 \partial_\nu\pi_\beta \partial^\nu\pi^\beta \nn\\
&&\qquad\quad
-\frac{1}{32}\frac{f^4}{16\pi^2}\partial_\mu\pi_\alpha \partial^\mu\pi^\alpha\left( 1-\frac{1}{2}\log\frac{-\square}{\mu^2}\right) \partial_\nu\pi_\beta \partial^\nu\pi^\beta
\nn\\
&&\qquad\quad
-\frac{1}{32}\frac{f^4}{16\pi^2}\partial_\mu\pi^\alpha\partial^\mu\pi _\beta\left( 1-\frac{1}{2}\log\frac{-\square}{\mu^2}\right) \partial_\nu\pi^\beta \partial^\nu\pi_\alpha
\nn\\
&&\qquad\quad
-\frac{1}{16}\frac{f^4}{16\pi^2}\partial^\mu\pi^\beta \partial_\nu\pi_\alpha \left( \frac{2}{9}-\frac{1}{12}\log\frac{-\square}{\mu^2}\right) \partial_\mu\pi^\alpha \partial^\nu\pi_\beta
\nn\\
&&\qquad\quad
+\frac{1}{16}\frac{f^4}{16\pi^2}\partial^\mu\pi_\alpha \partial_\nu\pi^\beta \left( \frac{4}{9}-\frac{1}{6}\log\frac{-\square}{\mu^2}\right) \partial_\mu\pi^\alpha \partial^\nu\pi_\beta
 +O(\pi^6)\Biggr\} \,,
\eea
where we have eliminated the subscripts from the couplings.
The renormalized couplings $f$, $\ell_1$ and $\ell_2$ have to be measured experimentally,
but the nonlocal terms are then completely determined and constitute therefore
low energy predictions of the theory.

These nonlocal terms enter in the computation of the Goldstone boson scattering amplitude $A(\pi_\alpha\pi_\beta\to \pi_\sigma\pi_\rho)=A(s,t,u)\delta_{\alpha\beta}\delta_{\sigma\rho}+A(t,s,u)\delta_{\alpha\sigma}\delta_{\beta\rho}+A(u,t,s)\delta_{\alpha\rho}
\delta_{\beta\sigma}$, which is obtained by taking four functional derivatives of the effective action in \eq{renactionnlsm} with respect to $\pi^\alpha$ after  performing the analytic continuation to Minkowski space and evaluating the expression at the particles external momenta. We get
\bea
A(s,t,u;\mu)&=&\frac{f^2}{4} s + \frac{\ell_1}{2}f^4 (t^2+u^2) 
+\ell_2 f^4s^2
\nn\\
&&\!\!\!\!\!\!\!\!\!-\frac{1}{3}\frac{f^4}{512\pi^2}\left( 2s^2\log\frac{-s}{\mu^2}+t(t-u)\log\frac{-t}{\mu^2}+u(u-t)\log\frac{-u}{\mu^2}\right)\,,
\eea
where $s=(p_1+p_2)^2$, $t=(p_1+p_3)^2$ and $u=(p_1+p_4)^2$
and all momenta are taken to be incoming. This result is well-known in the literature on chiral models \cite{gl}.

%\newpage

%%%%%%%%%%%%%%%%%%%%%%%%%%%%%%%%%%%
\section{Gravity with scalar field}
%%%%%%%%%%%%%%%%%%%%%%%%%%%%%%%%%%%

In this section we will consider another example of effective action for
an effective field theory, namely a scalar coupled minimally to gravity.
As first argued in \cite{donoghue1}
and confirmed by explicit calculations \cite{donoghue2,kk}
low energy gravitational scattering amplitudes can be calculated 
unambiguously in this effective field theory,
in spite of its perturbative non-renormalizability.
The reason is that low energy effects correspond to nonlocal terms 
in the effective action, that are non-analytic in momentum. 
Such terms are not affected by the divergences,
which manifest themselves as terms analytic in momentum.
Here we shall follow the logic of previous chapters and derive the terms in
the effective action containing up to two powers of curvature,
by integrating the flow equation.
There have been several calculations of divergences for a scalar coupled to gravity,
including also a generic potential \cite{bkk,takata,pietrykowski,steinwachs}.
As we shall see, some of these terms are related to the low energy ones
by simple properties, so that the two calculations partly overlap.
From a different point of view, the flow of scalar couplings due to gravity has
also been discussed in \cite{griguolo}, and with the aim of establishing
the existence of an UV fixed point, in \cite{narain1,narain2}.
Previous application of the FRGE to the calculation of some
terms in the gravitational EA have been given in
\cite{Codello:2010mj,Satz:2010uu,Codello:2011js,cdpp}.

%\subsection{Action}
\paragraph*{Action}

All calculations will be done in the Euclidean field theory.
We will study the flow of the EAA driven by an action of the form:
\be
\label{gravaction}
\Gamma_k[h,\bar C,C;g]=S_H[g+h]
+S_{m}[g+h,\phi]
+S_{gf}[h;g]
+S_{gh}[h,\bar C,C;g]\ ,
\ee
where $h_{\mu\nu}$ is the metric fluctuation, $\bar C^\mu$ and $C_\mu$
are anticommuting vector ghosts for diffeomorphisms,
$g_{\mu\nu}$ is the background metric.
The term
\be
S_H[g+h]=-\frac{1}{\kappa}\int d^{d}x\sqrt{\det(g+h)}R(g+h)\ ,
\ee
with $\kappa=16\pi G$, is the Hilbert action,
\be
S_{m}[g+h,\phi]= \int d^{d}x\sqrt{\det(g+h)}\left[\frac{1}{2}(g+h)^{\mu\nu}\partial_{\mu}\phi\partial_{\nu}\phi+V(\phi)\right]
\ee
is the matter action
\be
S_{gf}[h;g]=\frac{1}{2}\int d^{d}x\sqrt{g}\chi_{\mu}\chi^{\mu}\ ,
\mathrm{with}\ \
\chi_{\mu}=\nabla^{\nu}h_{\mu\nu}-\frac{1}{2}\nabla_{\mu}h\ ,
\ee
is the Feynman-de Donder-type gauge-fixing term 
(the gauge parameter $\alpha$ is set to one).
We will only need the ghost action for $h=0$, in which case it has the form
\be
S_{gh}[0,\bar C,C;g]=\int d^{d}x\sqrt{g}\,\bar{C}_{\mu}\left(-\square\delta_{\nu}^{\mu}-R_{\nu}^{\mu}\right)C^{\nu}\ .
\ee
The covariant derivative $\nabla$ and the curvature $R$ are constructed with the background metric
and we denote $\square=\nabla_{\mu}\nabla^{\mu}$.

Using equation (\ref{oneloopERGE}), this action generates all possible diffeomorphism-invariant
terms. We retain only those that are quadratic in ''curvatures'', where we include among curvatures
the Riemann tensor and its contractions, terms with two derivatives acting on one or two scalar fields
the potential and its derivatives.
\footnote{To some extent this parallels the treatment of mass terms in chiral
perturbation theory.}
Within this class of EAA's we can calculate the RG flow and integrate it analytically
to obtain an EA, which consists both of local and non-local terms.
In a more accurate treatment these new terms would all contribute to the
r.h.s. of the flow equation, but such calculations would be far more involved.

%\subsection{Hessian}
\paragraph{Hessian}

Arranging the fluctuation fields $h_{\mu\nu}$, $\delta\phi$
in a $d(d+1)/2+1$-dimensional column vector $\delta\Phi$, 
the total quadratic part can be written as:
\be
\frac{1}{2\kappa}\int d^{d}x\sqrt{g}\,
\delta\Phi^T \mathbb{H}\, \delta\Phi
=
\frac{1}{2\kappa}\int d^{d}x\sqrt{g}\,
\left(\begin{array}{cc}h_{\alpha\beta} & \delta\phi\end{array}\right)
\left(\begin{array}{c c} \mathbb{H}^{\alpha\beta\mu\nu}& \mathbb{H}^{\alpha\beta\,\cdot}\\ \mathbb{H}^{\,\cdot\,\mu\nu}&\mathbb{H}^{\,\cdot\,\cdot}\end{array}\right)
\left(\begin{array}{c} h_{\mu\nu}\\ \delta\phi\end{array}\right)
\ee
where the dot refers to the components in the space spanned by $\delta\phi$.
The Hessian $\mathbb{H}$ has the form
\be
\mathbb{H}=\mathbb{K}(-\square)+2\mathbb{V}^{\delta}\nabla_{\delta}+\mathbb{U}\ .
\ee
The coefficient of the second-derivative term is a quadratic form in field space
\be
\mathbb{K}=\left(\begin{array}{cc}
K^{\alpha\beta\mu\nu} & 0\\
0 & \kappa
\end{array}\right)
\ee
where
\be
K^{\alpha\beta\mu\nu}= \frac{1}{2}\left(\delta^{\alpha\beta,\mu\nu}-\frac{1}{2}g^{\alpha\beta}g^{\mu\nu}\right)
=\frac{1}{4}\left(g^{\mu\alpha}g^{\nu\beta}+g^{\nu\alpha}g^{\mu\beta}-g^{\alpha\beta}g^{\mu\nu}\right)
\ee
is the DeWitt metric.
(We denote $\delta_{\alpha\beta}{}^{\mu\nu}$ the identity in the space of symmetric tensors.)
Furthermore,
\begin{eqnarray}
\mathbb{V}^{\delta} & = & \left(\begin{array}{cc}
0 & -\kappa K^{\alpha\beta}{}^{\gamma\delta}\nabla_{\gamma}\phi\\
\kappa K^{\mu\nu\,\gamma\delta}\nabla_{\gamma}\phi & 0
\end{array}\right)
\\
\mathbb{U} & = & \left(\begin{array}{cc}
U^{\alpha\beta\mu\nu} 
& \frac{1}{2}\kappa g^{\alpha\beta}V'(\phi)\\
2\kappa K^{\mu\nu\gamma\delta}\nabla_{\gamma}\nabla_{\delta}\phi
+\frac{1}{2}\kappa g^{\mu\nu}V'(\phi) 
& \kappa V''(\phi)
\end{array}\right)\,,
\end{eqnarray}
with
\bea
U^{\alpha\beta\mu\nu} & = & 
K^{\alpha\beta\mu\nu}R
+\frac{1}{2}(g^{\mu\nu}R^{\alpha\beta}
+R^{\mu\nu}g^{\alpha\beta})
-\frac{1}{4}\left(g^{\alpha\mu}R^{\beta\nu}
+g^{\alpha\nu}R^{\beta\mu}
+g^{\beta\mu}R^{\alpha\nu}
+g^{\beta\nu}R^{\alpha\mu}\right)
\nonumber\\
&&
\!\!\!\!\!\!\!\!
-\frac{1}{2}(R^{\alpha\mu\beta\nu}+R^{\alpha\nu\beta\mu})
+\kappa\Bigg[
-\frac{1}{2}K^{\alpha\beta\mu\nu}(\nabla\phi)^2
%-V\delta_{\alpha\beta}{}^{\mu\nu}
-\frac{1}{4}(g^{\alpha\beta}\nabla^\mu\phi\nabla^\nu\phi
+g^{\mu\nu}\nabla^\alpha\phi\nabla^\beta\phi)
\\
&&
\!\!\!\!\!\!\!\!
+\frac{1}{4}\left(g^{\alpha\mu}\nabla^{\beta}\phi\nabla^\nu\phi
+g^{\alpha\nu}\nabla^{\beta}\phi\nabla^\mu\phi
+g^{\beta\mu}\nabla^{\alpha}\phi\nabla^\nu\phi
+g^{\beta\nu}\nabla^{\alpha}\phi\nabla^\mu\phi\right)
-K^{\alpha\beta\mu\nu}V\Bigg]\ .
\nonumber
\eea
It is convenient to extract an overall factor of $\mathbb{K}$
and write the Hessian as $\mathbb{H}=\mathbb{K}\Delta$ where
\be
\Delta=\mathbb{I}(-\square)+2\mathbb{Y}^{\delta}\nabla_{\delta}+\mathbb{W}
\ee
is a linear operator in field space and therefore has the index structure
\be
\Delta=
\left(\begin{array}{c c} \Delta_{\alpha\beta}{}^{\mu\nu}& \Delta_{\alpha\beta}{}^{\cdot}\\ \Delta_{\,\cdot}{}^{\mu\nu}&\Delta_{\,\cdot}{}^{\cdot}\end{array}\right)\ .
\label{deltadefinition}
\ee
The coefficients in the operator $\Delta$ are related to those of the Hessian by
$\mathbb{Y}^\delta=\mathbb{K}^{-1}\mathbb{V}^{\delta}$\ ,
$\mathbb{W}=\mathbb{K}^{-1}\mathbb{U}$,
where
\be
K^{-1}_{\mu\nu}{}^{\alpha\beta}= 2\delta_{\mu\nu}{}^{\alpha\beta}-\frac{2}{d-2}g_{\mu\nu}g^{\alpha\beta}\ .
\ee
Note $\mathbb{W}$ need not to be symmetric,
in fact the question is not even well-posed because of the different
position of the indices.
Explicit calculation leads to the following expressions:
\be
\mathbb{Y}^{\delta}=
\left(\begin{array}{cc}
0 & -\kappa \delta_{\alpha\beta}{}^{\gamma\delta}\nabla_{\gamma}\phi
\\
K^{\mu\nu\,\gamma\delta}\nabla_{\gamma}\phi & 0
\end{array}\right)
\ee
\be
\mathbb{W}=
\left(\begin{array}{cc}
W_{\alpha\beta}{}^{\mu\nu}
& -\frac{2}{d-2}\kappa g_{\alpha\beta}V'(\phi)\\
2 K^{\mu\nu\,\gamma\delta}\nabla_{\gamma}\nabla_{\delta}\phi
+\frac{1}{2}g^{\mu\nu}V'(\phi) 
& V''(\phi)
\end{array}\right)\,,
\ee
where
\be
W_{\alpha\beta}{}^{\mu\nu} = 2\,U_{\alpha\beta}{}^{\mu\nu}
-\frac{d-4}{d-2}\,g_{\alpha\beta}\left[
R^{\mu\nu}-\frac{1}{2}R\,g^{\mu\nu}
-\frac{\kappa}{2}\left(\nabla^\mu\phi\nabla^\nu\phi
-\frac{1}{2}g^{\mu\nu}(\nabla\phi)^2\right)
\right]
-\frac{\kappa}{2}g_{\alpha\beta}g^{\mu\nu}V\ .
\ee
%Note that unlike in $\mathbb{V}$ and $\mathbb{U}$, the dimensions of the top and bottom rows of $\mathbb{Y}$ and $\mathbb{W}$ 
%are different.

\paragraph{Completing the square}

In order to use the standard heat kernel formulas for minimal Laplace-type operators,
we have to eliminate the first order terms $\mathbb{Y}\cdot\nabla$.
This can be achieved by absorbing them in a redefinition of the covariant derivative:
$\tilde{\nabla}_{\mu}=\nabla_{\mu}\mathbb{I}-\mathbb{Y}_{\mu}$.
Then  $\Delta$ in (\ref{deltadefinition}) can be rewritten as
\begin{eqnarray}
\Delta & = & -\tilde{\nabla}_{\mu}\tilde{\nabla}^{\mu}+\tilde\mathbb{W}\\
\tilde\mathbb{W} & = & \mathbb{W}-\nabla_{\mu}\mathbb{Y}^{\mu}
+\mathbb{Y}_{\mu}\mathbb{Y}^{\mu}\,.
\end{eqnarray}
To compute $\tilde{\mathbb{W}}$ we need the following intermediate results:
\be
\nabla_\mu\mathbb{Y}^\mu=
\left(\begin{array}{cc}
0 & -\kappa\nabla_{\alpha}\nabla_{\beta}\phi\\
K^{\mu\nu\,\gamma\delta}\nabla_{\delta}\nabla_{\gamma}\phi & 0
\end{array}\right)
\ \
\mathbb{Y}_\mu\mathbb{Y}^\mu= 
\left(\begin{array}{cc}
-\kappa\delta_{\alpha\beta}{}^{\gamma\delta}K^{\mu\nu\,\epsilon}{}_\delta\nabla_{\gamma}\phi\nabla_{\epsilon}\phi & 0\\
0 & -\kappa(\nabla\phi)^{2}
\end{array}\right)
\ee
and
\be
K_{\gamma\delta}{}^{\epsilon\delta}=\frac{d}{4}\delta_{\gamma}^{\epsilon}
\qquad
\delta_{\alpha\beta}{}^{\gamma\delta}
K^{\mu\nu\,\epsilon}{}_\delta\nabla_{\gamma}\phi\nabla_{\epsilon}\phi
=-\frac{1}{2}\delta_{(\alpha}^{(\mu}\nabla_{\beta)}\phi\nabla^{\nu)}\phi
+\frac{1}{4}g^{\mu\nu}\nabla_{\alpha}\phi\nabla_{\beta}\phi\,.
\ee
Collecting, we find
\be
\tilde{\mathbb{W}}=\left(\begin{array}{cc}
A_{\alpha\beta}{}^{\mu\nu} & B_{\alpha\beta}\\
C^{\mu\nu} & D
\end{array}\right)
\ee
with:
\begin{eqnarray*}
A_{\alpha\beta}{}^{\mu\nu} & = & W_{\alpha\beta}{}^{\mu\nu}
-\frac{1}{2}\kappa\delta_{(\alpha}^{(\mu}\nabla_{\beta)}\phi\nabla^{\nu)}\phi
+\frac{1}{4}\kappa g^{\mu\nu}\nabla_{\alpha}\phi\nabla_{\beta}\phi
%-\delta_{\alpha\beta}{}^{\mu\nu}V
\\
B_{\alpha\beta} & = & -\frac{2}{d-2}\kappa g_{\alpha\beta}V'(\phi)
+\kappa \nabla_{\alpha}\nabla_{\beta}\phi\\
C^{\mu\nu} & = &  K^{\mu\nu\,\alpha\beta}\nabla_{\alpha}\nabla_{\beta}\phi
+\frac{1}{2} g^{\mu\nu}V'(\phi)\\
D & = & -\kappa(\nabla\phi)^{2}+V''(\phi)\,.
\end{eqnarray*}
The curvature of covariant derivatives, 
$\tilde\Omega_{\mu\nu}=[\tilde\nabla_\mu,\tilde\nabla_\nu]$, 
is related to the curvature of the original covariant derivative, 
$\Omega=[\nabla_\mu,\nabla_\nu]$ by:
\be
\tilde{\Omega}_{\mu\nu}=\Omega_{\mu\nu}-\nabla_{\mu}\mathbb{Y}_{\nu}
+\nabla_{\nu}\mathbb{Y}_{\mu}
+\mathbb{Y}_{\mu}\mathbb{Y}_{\nu}-\mathbb{Y}_{\nu}\mathbb{Y}_{\mu}\,.
\ee
To compute $\tilde{\Omega}_{\kappa\lambda}$ we need:
\be
2\nabla_{[\kappa}\mathbb{Y}_{\lambda]}=\left(\begin{array}{cc}
0 & -2\kappa\delta_{\alpha\beta}{}^{\gamma\delta}g_{[\lambda|\delta|}
\nabla_{\kappa]}\nabla_{\gamma}\phi
\\
2K^{\mu\nu\,\gamma\delta}g_{[\lambda|\delta|}\nabla_{\kappa]}\nabla_{\gamma}\phi & 0
\end{array}\right)
\ee
and
\be
2\mathbb{Y}_{[\kappa}\mathbb{Y}_{\lambda]}=\left(\begin{array}{cc}
-2\kappa\delta_{\alpha\beta\,\gamma[\kappa}Z_{|\epsilon|\lambda]}^{\mu\nu}\nabla^{\gamma}\phi\nabla^{\epsilon}\phi & 0\\
0 & 0 
\end{array}\right)\,.
\ee
Then we find:
\be
\tilde{\Omega}_{\kappa\lambda} = 
\left(\begin{array}{cc}
(\tilde{\Omega}_{\kappa\lambda})_{\alpha\beta}{}^{\mu\nu} & (\tilde{\Omega}_{\kappa\lambda})_{\alpha\beta}{}^{\,\cdot}\\
(\tilde{\Omega}_{\kappa\lambda})_{\,\cdot}{}^{\mu\nu} & 0
\end{array}\right)\,,
\ee
where:
\begin{eqnarray*}
(\tilde{\Omega}_{\kappa\lambda})_{\alpha\beta}{}^{\mu\nu} & = & (\Omega_{\kappa\lambda})_{\alpha\beta}{}^{\mu\nu}
-2\kappa\delta_{\alpha\beta\,\gamma[\kappa|}K_{\epsilon|\lambda]}{}^{\mu\nu}\nabla^{\gamma}\phi\nabla^{\epsilon}\phi
\\
(\tilde{\Omega}_{\kappa\lambda})_{\alpha\beta}{}^{\,\cdot} & = & 2\kappa\delta_{\alpha\beta}{}^{\gamma\delta}g_{[\lambda|\delta}\nabla_{|\kappa]}\nabla_{\gamma}\phi
\\
(\tilde{\Omega}_{\kappa\lambda})_{\,\cdot}{}^{\mu\nu} & = & 
-2 K^{\mu\nu\,\gamma\delta}g_{[\lambda|\delta}\nabla_{|\kappa]}\nabla_{\gamma}\phi\,.
\end{eqnarray*}

%\subsection{Heat kernel coefficients}
\paragraph{Heat kernel coefficients}

We compute first the local heat kernel coefficients of the operator $\Delta$, using (\ref{hkcoefficient}).
The first two are
\be
\label{blowdelta}
\tr\mathbf{b}_0(\Delta)=11\ ,\qquad 
\tr\mathbf{b}_2(\Delta)=-\frac{25}{6}R+2\kappa((\nabla\phi)^2)^2-V''+10\kappa V\,.
\ee
For the calculation of $\mathbf{b}_4$ we need a few preliminary results.
Using that $\tr\mathbb{I}=11$ and defining $\mathbb{P}=\tilde\mathbb{W}-\frac{R}{6}\mathbb{I}$, the heat kernel coefficient (\ref{hkcoefficient}) can be rewritten, in four dimensions, 
in the more compact form
\be
\label{lesl}
\tr b_4(\Delta)=
\frac{11}{180}\left(\text{Riem}^{2}-\text{Ric}^{2}\right)
+\frac{1}{2}\textrm{tr}\,\mathbb{P}^2+\frac{1}{12}\textrm{tr}\tilde\Omega^{2}\ .
\ee
In the evaluation of the last two terms we use the following traces:
\begin{eqnarray*}
\textrm{tr}\,\mathbb{P}^2&=&
\tr\tilde\mathbb{W}^2-\frac{1}{3}R\,\tr\tilde\mathbb{W}+\frac{11}{36}R^2\ ,
\\
\textrm{tr}\,\tilde{\mathbb{W}}&=&A_{\mu\nu}{}^{\mu\nu}+D\ ,
\\
A_{\mu\nu}{}^{\mu\nu}&=&6R-\kappa(\nabla\phi)^2-10\kappa V\ ,
\\
\textrm{tr}\,\tilde{\mathbb{W}}^2 & = & 
A_{\mu\nu}{}^{\alpha\beta}A_{\alpha\beta}{}^{\mu\nu}+2B_{\alpha\beta}C^{\alpha\beta}+D^{2}
\\
A_{\mu\nu}{}^{\alpha\beta}A_{\alpha\beta}{}^{\mu\nu}&=&
3R_{\mu\nu\rho\sigma}R^{\mu\nu\rho\sigma}
-6R_{\mu\nu}R^{\mu\nu}+5R^2
-\frac{3}{2}\kappa R(\nabla\phi)^2
-12\kappa V\,R
\\
&&
+\frac{7}{4}\kappa^2((\nabla\phi)^2)^2
+2\kappa^2 V(\nabla\phi)^2
+10\kappa^2V^2
\\
B_{\alpha\beta}C^{\alpha\beta}&=&
\kappa V'\nabla^2\phi
-2\kappa V^{\prime 2}
+\frac{1}{2}\kappa \nabla_\mu\nabla_\nu\phi\nabla^\mu\nabla^\nu\phi
-\frac{1}{4}\kappa(\nabla^2\phi)^2
\\
\frac{1}{2}\tr\mathbb{P}^2 & = & \frac{3}{2}\text{Riem}^{2}-3\text{Ric}^{2}+\frac{119}{72}R^{2}\\
 &  & +\frac{11}{8}\kappa^2\nabla_{\alpha}\phi\nabla^{\alpha}\phi\nabla_{\beta}\phi\nabla^{\beta}\phi-\frac{1}{4}\kappa\nabla^{2}\phi\nabla^{2}\phi+\frac{1}{2}\kappa\nabla_{\beta}\nabla_{\alpha}\phi\nabla^{\beta}\nabla^{\alpha}\phi\\
 &  & +\kappa\left(\kappa V(\phi)-\frac{5}{12}R-V''(\phi)\right)\nabla_{\alpha}\phi\nabla^{\alpha}\phi\\
 &  & +\kappa V'(\phi)\nabla^{2}\phi+5\kappa^2V^{2}(\phi)-\frac{13}{3}\kappa R\, V(\phi)\\
 &  & -2\kappa V'(\phi)^{2}-\frac{R}{6}V''(\phi)+\frac{V''(\phi)^2}{2}\,.
\end{eqnarray*}

For the last term in (\ref{lesl}) we need
\be
\textrm{tr}\,\tilde{\Omega}_{\mu\nu}\tilde{\Omega}^{\mu\nu}= (\tilde{\Omega}_{\mu\nu})_{\alpha\beta}{}^{\gamma\delta}
(\tilde{\Omega}^{\mu\nu})_{\gamma\delta}{}^{\alpha\beta}
+2(\tilde{\Omega}_{\mu\nu})_{\alpha\beta}(\tilde{\Omega}^{\mu\nu})^{\alpha\beta}\,,
\ee

where
\begin{eqnarray*}
(\tilde{\Omega}_{\mu\nu})_{\alpha\beta}{}^{\gamma\delta}
(\tilde{\Omega}^{\mu\nu})_{\gamma\delta}{}^{\alpha\beta} & = & -6R_{\mu\nu\rho\sigma}R^{\mu\nu\rho\sigma}+\kappa R(\nabla\phi)^{2}
+2\kappa R_{\mu\nu}\nabla^{\mu}\phi\nabla^{\nu}\phi
-\frac{3}{2}\kappa((\nabla\phi)^2)^2
\\
2(\tilde{\Omega}_{\mu\nu})_{\alpha\beta}(\tilde{\Omega}^{\mu\nu})^{\alpha\beta} & = & \kappa\left((\nabla^2\phi)^{2}-4\nabla_{\mu}\nabla_{\nu}\phi\nabla^{\mu}\nabla^{\nu}\phi\right)\,.
\end{eqnarray*}

We thus arrive at
\begin{eqnarray*}
\tr \mathbf{b}_{4}(\Delta) & = & \frac{191}{180}\text{Riem}^{2}
-\frac{551}{180}\text{Ric}^{2}
+\frac{119}{72}R^{2}
\\
 &  & 
+\frac{5}{4}\kappa^2((\nabla\phi)^2)^2
-\frac{1}{6}\kappa\nabla^{2}\phi\nabla^{2}\phi
+\frac{1}{6}\kappa\nabla_{\beta}\nabla_{\alpha}\phi\nabla^{\beta}\nabla^{\alpha}\phi
+\frac{1}{6}\kappa R_{\alpha\beta}\nabla^{\alpha}\phi\nabla^{\beta}\phi
\\
 &  & 
+\kappa\left(\kappa V(\phi)-\frac{1}{3}R-V''(\phi)\right)(\nabla\phi)^2
+\kappa V'(\phi)\nabla^{2}\phi
 \\
 &  & +5\kappa^2V^{2}(\phi)-2\kappa V'(\phi)^{2}-\frac{13}{3}\kappa R\, V(\phi)-\frac{R}{6}V''(\phi)+\frac{V''(\phi)^2}{2}\,.
\end{eqnarray*}
Using the relations
\be
\frac{1}{6}R_{\alpha\beta}\nabla^{\alpha}\phi\nabla^{\beta}\phi
-\frac{1}{6}\nabla^{2}\phi\nabla^{2}\phi
+\frac{1}{6}\nabla_{\beta}\nabla_{\alpha}\phi\nabla^{\beta}\nabla^{\alpha}\phi=0
\ee
\be
-V''(\phi)(\nabla\phi)^2+V'(\phi)\nabla^2\phi=-2V''(\phi)(\nabla\phi)^2
\ee
\be
\label{euler}
R_{\mu\nu\rho\sigma}R^{\mu\nu\rho\sigma}=
4R_{\mu\nu}R^{\mu\nu}-R^2\,,
\ee
which hold under an integral modulo surface terms, one can rewrite
\begin{eqnarray}
\tr \mathbf{b}_{4}(\Delta) & = & %\frac{191}{180}\text{Riem}^{2}
\frac{71}{60}\text{Ric}^{2}+\frac{71}{120}R^{2}
-\frac{1}{3}\kappa R(\nabla\phi)^2
-\frac{13}{3}\kappa R\, V(\phi)
-\frac{1}{6}RV''(\phi)
+\frac{5}{4}\kappa^2((\nabla\phi)^2)^2
\nonumber\\
 &  & 
+\kappa^2V(\phi)(\nabla\phi)^2
-2\kappa V''(\phi)(\nabla\phi)^2
+5\kappa^2V^{2}(\phi)-2\kappa V'(\phi)^{2}
+\frac{V''(\phi)^2}{2}\ .
\label{bfourdelta}
\end{eqnarray}
The nonlocal heat kernel can be computed from equation (\ref{nlhk}).
Using again the traces given above, the coefficient of $s^2$ is found to be
\begin{eqnarray}
&&
R_{\mu\nu}(11f_{Ric}+6f_U-24f_\Omega)R^{\mu\nu}
+R(11f_R+6f_{RU}+2f_U+6f_\Omega)R
\nonumber\\
&&
+\kappa R\left(-2f_{RU}-\frac{3}{2}f_U+f_\Omega\right)(\nabla\phi)^2
+2\kappa R^{\mu\nu}f_\Omega \nabla_\mu\phi\nabla_\nu\phi
+R f_{RU}V''
+\kappa R(-10f_{RU}-12f_U)V
\nonumber\\
&&
+\kappa\nabla_\mu\nabla_\nu\phi(f_U-4f_\Omega)\nabla^\mu\nabla^\nu\phi
+\kappa\nabla^2\phi\left(-\frac{1}{2}f_U+f_\Omega\right)\nabla^2\phi
+2\kappa\nabla^2\phi f_U V'
-2\kappa(\nabla\phi)^2 f_U V''
\nonumber\\
&&
+2\kappa^2 V f_U (\nabla\phi)^2
-4\kappa V' f_U V'
+V'' f_U V''
+\left(\frac{11}{4}f_{U}-\frac{3}{2}f_{\Omega}\right)\kappa^{2}\left(\left(\nabla\phi\right)^{2}\right)^{2}
+10\kappa^2 V f_U V\,.
\end{eqnarray}
Using the expansion (\ref{sfexp}) one can check that the local part of this expression
agrees with (\ref{bfourdelta}).

One can compute in a similar way the heat kernel of the ghost operator
$(\Delta_{gh})_\mu^\nu=-\nabla^2\delta_\mu^\nu-R_\mu^\nu$.
We find, in four dimensions, the local heat kernel coefficients
\be
\tr\mathbf{b}_0(\Delta_{gh})=4\  ;\qquad\ 
\tr\mathbf{b}_2(\Delta_{gh})=-\frac{1}{3}R\ ;\qquad
\tr\mathbf{b}_4(\Delta_{gh})=\frac{7}{30}R_{\mu\nu}R^{\mu\nu}+\frac{17}{60}R^2\ ,
\label{bfourdeltagh}
\ee
where we used again (\ref{euler}).
The coefficient of $s^2$ in (\ref{nlhk}) is
\be
R_{\mu\nu}(4f_{Ric}
+f_U-4f_\Omega)R^{\mu\nu}
+R(4f_R-f_{RU}+f_\Omega)R\ .
\ee
We are now ready to write the flow equation.

%\subsection{Flow equations}
\paragraph{Flow equations}

We write the one-loop flow equation for the ``single metric'' bEAA $\Gamma_k[g]\equiv\Gamma_k[0,0,0;g]$.
It consists of two terms:
\begin{equation}
\partial_{t}{\Gamma}_{k}[g]=\frac{1}{2}\textrm{Tr}\frac{\partial_{t}R_{k}(\Delta)}{\Delta+R_{k}(\Delta)}-\textrm{\textrm{Tr}}\frac{\partial_{t}R_{k}(\Delta_{gh})}{\Delta_{gh}+R_{k}(\Delta_{gh})}\ .\label{3}
\end{equation}
The first comes from the graviton and scalar fluctuations, the second from the ghosts.
Using equation (\ref{adam}) and the heat kernel discussed in the previous section,
the flow equation for the bEAA is:
\begin{eqnarray}
\partial_t\Gamma_k[g]&=&
\frac{1}{32\pi^2}\int d^4x\sqrt{g}\Bigg\{
3k^4
+k^2\left(
-\frac{29}{3}R
+4\kappa(\nabla\phi)^2
-2 V''
+20\kappa V\right)
\nonumber\\
&&
\qquad\qquad\qquad
+R_{\mu\nu}\,g_1(\tilde z)\,R^{\mu\nu}
+R g_2(\tilde z)R
\nonumber\\
&&
\qquad\qquad\qquad
+\kappa R g_3(\tilde z)(\nabla\phi)^2
+\kappa R^{\mu\nu}g_4(\tilde z) \nabla_\mu\phi\nabla_\nu\phi
+R g_5(\tilde z)V''
+\kappa R g_6(\tilde z) V
\nonumber\\
&&
\qquad\qquad\qquad
+\kappa\nabla_\mu\nabla_\nu\phi\, g_7(\tilde z)\nabla^\mu\nabla^\nu\phi
+\kappa\nabla^2\phi\, g_8(\tilde z)\nabla^2\phi
+\kappa^2(\nabla\phi)^2 g_9(\tilde z) (\nabla\phi)^2
\nonumber\\
&&
\qquad\qquad\qquad
+\kappa\nabla^2\phi\, g_{10}(\tilde z) V'
-\kappa(\nabla\phi)^2 g_{10}(\tilde z) V''
+\kappa^2 (\nabla\phi)^2 g_{10}(\tilde z) V
\nonumber\\
&&
\qquad\qquad\qquad
-2\kappa V' g_{10}(\tilde z) V'
+\frac{1}{2}V'' g_{10}(\tilde z) V''
+5\kappa^2 V g_{10}(\tilde z) V
\Bigg\}\,,
\label{gravflowgen}
\end{eqnarray}
where the functions $g_a$, $a=1\ldots 10$ are linear combinations of
the functions $g_{Ric}$, $g_R$, $g_U$, $g_{RU}$, $g_\Omega$ 
given in the second column of the following Table:
\\
\begin{center}
    \begin{tabular}{|r |l | c | c | c |}
    \hline
     $a$ &\quad\quad $g_a$ & $A_a$ & $B_a$ & $C_a$ \\
      \hline
   1 &\quad  $3g_{Ric}+4g_U-16g_\Omega$ & $\frac{43}{30}$  & $-\frac{148}{15}$ & $-\frac{8}{5}$ \\ \hline
   2 &\quad $3g_R+2g_U+8g_{RU}+4g_\Omega$ & $\frac{1}{20}$  & $\frac{37}{5}$ & $\frac{1}{5}$ \\ \hline
  3 &\quad $-2g_{RU}-\frac{3}{2}g_U+g_\Omega$ & $-\frac{2}{3}$  & $-\frac{2}{3}$ & $0$ \\ \hline
  4 &\quad $2 g_\Omega$ & $\frac{1}{3}$  & $\frac{4}{3}$ & $0$ \\ \hline
  5&\quad $g_{RU}$  & $-\frac{1}{3}$  & $\frac{2}{3}$ & $0$ \\ \hline
  6 &\quad $-10g_{RU}-12g_U$ & $-\frac{26}{3}$  & $-\frac{20}{3}$ & $0$ \\ \hline
  7 &\quad $g_U-4g_\Omega$ & $\frac{1}{3}$  & $-\frac{8}{3}$ & $0$ \\ \hline
  8 &\quad $-\frac{1}{2}g_U+g_\Omega$ & $-\frac{1}{3}$  & $\frac{2}{3}$ & $0$ \\ \hline
  9 &\quad $\frac{11}{4}g_U-\frac{3}{2}g_\Omega$ & $\frac{5}{2}$  & $-1$ & $0$ \\ \hline
  10 &\quad $2 g_U$ & $2$  & $0$ & $0$ \\ \hline
    \end{tabular}
\end{center}
Note that only the first two receive contributions from the ghosts.
Next we recall that using the optimized cutoff, the functions $g_{Ric}$, \ldots $g_\Omega$ are given
by equations (\ref{eve1}-\ref{eve5}).
In the present case there is no dependence on $\omega$, so dropping this argument
the functions $g_a(\tilde z)=g_a(z,k)$ can all be written in the form:
\begin{equation}
\label{strf}
g_a(\tilde z)=A_a+\left(-A_a+\frac{B_a}{\tilde z}+\frac{C_a}{\tilde z^{2}}\right)
\sqrt{1-\frac{4}{\tilde z}}\,\theta(\tilde z-4)\ ,
\end{equation}
where the coefficients $A_a$, $B_a$ and $C_a$ are given in the remaining columns of the Table above.

%\subsection{Integration of the flow}
\paragraph{Integration of the flow}

We have found that the classical action (\ref{gravaction}) generates the flow (\ref{gravflowgen}).
The flow is finite but integrating over $k$ up to some scale $\Lambda$ and taking the limit $\Lambda\to\infty$
it generates divergences in the effective action.
The first line is proportional to terms that are already present in the action
and, upon integration over $k$, corresponds to quartic and quadratic divergences.
The remaining terms were not originally present in the action.
Equation (\ref{strf}) allows us to split these terms into local parts (the first term)
and nonlocal parts (the rest).
The local parts correspond to logarithmic UV divergences.
The nonlocal parts vanish for $k^2>z/4$, signalling its infrared character.

The new terms in the flow equation force us to consider a more general class
of EAA's that, in addition to the terms that were originally present in (\ref{gravaction})
also contains for each $g_a(\tilde z)\equiv g_a(z,k)$, $a=1,\ldots,10$ in (\ref{gravflowgen}),  
a corresponding form factor $f_{a,k}(z)$.
In principle all the terms in the last two lines of (\ref{gravflowgen}) could appear 
with different couplings,
but in the present truncation they all renormalize in the same way.
The functions $g_a$ are the beta functions of the form-factors $f_a$:
\be
\partial_t f_{a,k}(z)=\frac{1}{32\pi^2}g_a(\tilde z)\ .
\ee

Even though our main interest is in infrared physics, it is useful to consider first the
divergent part of the action.
For a fixed $z$ and for $k^2>z/4$, the step function in (\ref{strf}) vanishes.
In this case the integration of the flow only gives power and logaritmic terms.
We find for large $\Lambda$:
\begin{eqnarray}
\Gamma_\Lambda[g]&=&
\frac{1}{32\pi^2}\int d^4x\sqrt{g}\Bigg\{
3\Lambda^4
+\left(-\frac{29}{3}R
+4\kappa(\nabla\phi)^2
-2 V''
+20\kappa V\right)\Lambda^2
\nonumber\\
&&
\qquad%\qquad
+\Bigg(\frac{43}{30}R_{\mu\nu}R^{\mu\nu}
+\frac{1}{20}R^2
-\frac{2}{3}\kappa R (\nabla\phi)^2
-\frac{26}{3}\kappa R\, V(\phi)
-\frac{1}{3}RV''(\phi)
\nonumber\\
 &  & \qquad%\qquad
+\frac{5}{2}\kappa^2((\nabla\phi)^2)^2
+2\kappa^2V(\phi)(\nabla\phi)^2
-4\kappa V''(\phi)(\nabla\phi)^2
\nonumber\\
 &  & \qquad%\qquad
+10\kappa^2V^{2}(\phi)
-4\kappa V'(\phi)^{2}
+V''(\phi)^2\Bigg)
\log\left(\frac{\Lambda^2}{\mu^2}\right)
+\mathrm{finite\ terms}
\Bigg\}\,,
\label{divergent}
\end{eqnarray}
where we have introduced an arbitrary renormalization scale $\mu$.
The last three lines agree with the results of \cite{steinwachs},
specialized to a single component field.
The finite terms have the same structure of the ones that are written 
but with $\Lambda$-independent, finite, coefficients.
These correspond to the arbitrariness in the choice of the initial conditions for the flow.

We now integrate the flow equations from the UV scale $\Lambda$ down to zero. 
The first line in (\ref{gravflowgen}), which depends on powers of the cutoff scale $k$,
produces simply the terms in the first line of (\ref{divergent}).

Next we discuss the logarithmic terms.
For each form factor the integration is to be
performed keeping $z$ fixed. For a generic IR scale $k$ we have
\begin{equation}
f_{a,\,\Lambda}(z)-f_{a,\, k}(z)
=\frac{1}{32\pi^2}\int_{k}^{\Lambda}\frac{dk'}{k'}g_a\left(\frac{z}{k'^{2}}\right)
=\frac{1}{64\pi^2}\int_{z/\Lambda^2}^{z/k^2}
\frac{du}{u}\tilde g_a(u)\ ,
\end{equation}
where in the second step we changed variables of integration to $u=z/k^2$.
Now we insert the explicit form (\ref{strf}). 
We assume that $z/\Lambda^2\ll 4$ and $z/k^2\gg 4$.
The integration can be done explicitly and in the limit $k\to 0$ one finds
\begin{equation}
f_{a,0}(z)
=f_{a,\Lambda}(z)
-\frac{A_a}{64\pi^2}\log\left(\frac{\Lambda^2}{\mu^2}\right)
+\frac{A_a}{64\pi^2}\log\left(\frac{z}{\mu^2}\right)
-\frac{1}{32\pi^2}\left(A_a+\frac{B_a}{12}+\frac{C_a}{120}\right)\ .
\end{equation}

Note the following fact: if we had just integrated the local part of the form factor,
namely the first term in (\ref{strf}), from $k$ to $\Lambda$,
we would have obtained $\frac{A}{32\pi^2}\log(\Lambda^2/k^2)$
which is both UV and IR divergent.
When we integrate the full form factor, for $k\to0$ in (\ref{strf}) $\tilde z$ becomes large, 
the theta function is one
and the square root tends to one, so that the $A$-terms cancel.
Thus, by integrating the full nonlocal form factor, we obtain
a result that is UV divergent but IR finite.

While the nonlocal terms are finite and entirely unambiguous,
the local terms in the effective action $\Gamma_0$ are not fixed
and have to be determined by matching the form of the EA with experimental data.
This is achieved by means of renormalization conditions.
In the present formalism, these correspond to the choice of $\Gamma_\Lambda$.

The initial conditions for the form-factors can be chosen to eliminate the $\Lambda$-dependence
in the EA. This requires
\begin{equation}
f_{a,\Lambda}(z)=
\frac{A_a}{64\pi^2}\log\left(\frac{\Lambda^2}{\mu^2}\right)
+\gamma_a\,,
\end{equation}
where $\gamma_a$ are arbitrary constants, corresponding to the finite terms in (\ref{divergent}).
Then the form-factors in the EA are
\begin{equation}
\label{enzo}
f_{a,0}(z)
=\frac{A_a}{64\pi^2}\log\left(\frac{z}{\mu^2}\right)+\gamma_a-\frac{1}{32\pi^2}\left(A_a+\frac{B_a}{12}+\frac{C_a}{120}\right)\ ,
\end{equation}
and the EA has the form
\begin{eqnarray}
\label{leading_logs}
\Gamma_0[g]&=&
\frac{1}{32\pi^2}\int d^4x\sqrt{g}\Bigg\{
\frac{43}{30}R_{\mu\nu}\log\left(\frac{-\Box}{\mu^2}\right)\,R^{\mu\nu}
+\frac{1}{20}R \log\left(\frac{-\Box}{\mu^2}\right)R
\nonumber\\
&&
\qquad\qquad\qquad
-\frac{2}{3}\kappa R \log\left(\frac{-\Box}{\mu^2}\right)(\nabla\phi)^2
+\frac{1}{3}\kappa R^{\mu\nu}\log\left(\frac{-\Box}{\mu^2}\right) \nabla_\mu\phi\nabla_\nu\phi
\nonumber\\
&&
\qquad\qquad\qquad
-\frac{1}{3}R \log\left(\frac{-\Box}{\mu^2}\right)V''
-\frac{26}{3}\kappa R \log\left(\frac{-\Box}{\mu^2}\right) V
\nonumber\\
&&
\qquad\qquad\qquad
+\frac{1}{3}\kappa\nabla_\mu\nabla_\nu\phi\, \log\left(\frac{-\Box}{\mu^2}\right)\nabla^\mu\nabla^\nu\phi
-\frac{1}{3}\kappa\nabla^2\phi\, \log\left(\frac{-\Box}{\mu^2}\right)\nabla^2\phi
\nonumber\\
&&
\qquad\qquad\qquad
+\frac{5}{2}\kappa^2(\nabla\phi)^2 \log\left(\frac{-\Box}{\mu^2}\right) (\nabla\phi)^2
+2\kappa\nabla^2\phi\, \log\left(\frac{-\Box}{\mu^2}\right) V'
\nonumber\\
&&
\qquad\qquad\qquad
-2\kappa(\nabla\phi)^2 \log\left(\frac{-\Box}{\mu^2}\right) V''
+2\kappa^2 (\nabla\phi)^2\log\left(\frac{-\Box}{\mu^2}\right) V
\nonumber\\
&&
\qquad\qquad\qquad
+10\kappa^2 V \log\left(\frac{-\Box}{\mu^2}\right) V
-4\kappa V' \log\left(\frac{-\Box}{\mu^2}\right) V'
+V'' \log\left(\frac{-\Box}{\mu^2}\right) V''
\nonumber\\
&&
\qquad\qquad\qquad
+\mathrm{local\ terms}
\Bigg\}\ .
\end{eqnarray}
The local terms are operators of the form appearing in(\ref{divergent}), with arbitrary finite coefficients.
These coefficients are related to the ``bare'' couplings by renormalization conditions.
To discuss them it is more transparent to specify the form of the potential, e.g.
$$
V(\phi)=E+\frac{1}{2}m^2\phi^2
+\frac{1}{24}\phi^4\ .
$$
Recalling that all the couplings in the EAA are $k$-dependent,
we denote by subscripts $\Lambda$ and $0$ the ``bare'' and ``renormalized'' couplings respectively.
Then we find that the terms that are already present in the action are renormalized as follows:
\begin{eqnarray}
E_0&=&E_\Lambda
+\frac{1}{32\pi^2}\left[
3\Lambda^4
+(20\kappa E-2m^2)\Lambda^2
+(10\kappa^2 E^2+m^4)\log\left(\frac{\Lambda^2}{\mu^2}\right)
\right]
\\
\frac{1}{2}m^2_0&=&\frac{1}{2}m^2_\Lambda
+\frac{1}{32\pi^2}\left[(10m^2\kappa-\lambda)\Lambda^2
+(10\kappa^2 E m^2-4\kappa m^4+m^2\lambda)\log\left(\frac{\Lambda^2}{\mu^2}\right)
\right]
\\
\frac{1}{24}\lambda_0&=&\!\frac{1}{24}\lambda_\Lambda
+\frac{1}{32\pi^2}\left[
\frac{20}{24}\lambda\kappa\Lambda^2
\!+\!\left(\frac{5}{2}\kappa^2 m^4
+\frac {5}{6}\lambda\kappa^2 E
-\frac{2}{9}\lambda \kappa m^2
+\frac{1}{4}\lambda^2\right)\!\log\left(\frac{\Lambda^2}{\mu^2}\right)
\!\right]
\\
Z_0&=&1+
\frac{1}{32\pi^2}
\left[4\kappa\Lambda^2
+\left(2\kappa^2E-4\kappa m^2 \right)\log\left(\frac{\Lambda^2}{\mu^2}\right)\right]\ .
\end{eqnarray}
The renormalization of Newton's constant is given by
\be
\frac{1}{\kappa_0}=
\frac{1}{\kappa_\Lambda}
-\frac{1}{32\pi^2}\left[
\frac{29}{3}\Lambda^2
+\left(-\frac{26}{3}\kappa E
-\frac{1}{3} m^2
\right)\log\left(\frac{\Lambda^2}{\mu^2}\right)
\right]\ .
\ee
In addition the following terms, not initially present in the action, are generated
\be
\xi_2\phi^2 R+\xi_4\phi^4 R+\tau(\nabla\phi)^2 R+w(\nabla\phi)^4+\alpha R^2+\beta R_{\mu\nu}R^{\mu\nu}\,,
\ee
with effective couplings
\begin{eqnarray}
\xi_2&=&
\frac{1}{32\pi^2}
\left(-\frac{13}{3}\kappa m^2
-\frac{1}{6}\lambda \right)\log\left(\frac{\Lambda^2}{\mu^2}\right)
\label{andreas}
\\
\xi_4&=&
-\frac{1}{32\pi^2}
\frac{13}{36}\kappa \lambda
\log\left(\frac{\Lambda^2}{\mu^2}\right)
\\
\tau&=&
-\frac{1}{32\pi^2}
\frac{2}{3}\kappa
\log\left(\frac{\Lambda^2}{\mu^2}\right)
\\
w
&=&
\frac{1}{32\pi^2}
\frac{5}{2}\kappa^2
\log\left(\frac{\Lambda^2}{\mu^2}\right)
\\
\alpha&=&
\frac{1}{32\pi^2}
\frac{1}{20}\log\left(\frac{\Lambda^2}{\mu^2}\right)
\\
\beta&=&
\frac{1}{32\pi^2}
\frac{43}{30}\log\left(\frac{\Lambda^2}{\mu^2}\right)\,.
\end{eqnarray}
The last term in (\ref{andreas}) is due entirely to scalar loops and would also
be present if gravity was treated as an external field.
In this context this term has been discussed several times in the literature,
for example see \cite{stw}.

We note that all the renormalized couplings depend on the reference scale $\mu$.
This dependence encodes a different notion of renormalization group.
For the cosmological constant and Newton's constant this has been discussed e.g. in \cite{solashapiro}.

Finally let us comment on the finite local terms in (\ref{leading_logs}).
The choice 
$\gamma_a=\frac{1}{32\pi^2}\left(A_a+\frac{B_a}{12}+\frac{C_a}{120}\right)$
has the effect that all the local terms vanish.
The choice 
$\gamma_a=0$ leaves a residual finite term that can be easily calculated from (\ref{enzo}) and the Table.
For example the first term in the last line of (\ref{gravflowgen}) would leave in this case 
$\frac{1}{8\pi^2}\kappa m^4\phi^2$.
We see that the gravitational correction to the scalar mass is suppressed by the ratio $(m/M_{\rm Planck})^2$,
as one would naturally guess from the weakness of gravity at low energy.
In any case the constants $\gamma_a$ cannot be calculated
but have to be fixed by comparison to experiment.

From the point of view of the effective field theory approach,
the action (\ref{leading_logs}) contains part of the
terms needed to reconstruct the scalar-graviton vertex
in a perturbative expansion about flat space.
More precisely, it contains the terms with two generalized
curvatures, which correspond to Feynman diagrams
with a three-graviton and two-scalar-two-graviton vertices.
Other contributions corresponding to triangle diagrams
are encoded in terms with three generalized curvatures,
that we have not evaluated.
\bigskip

{\bf Acknowledgements}. This work is based in part on the
Ph.D theses presented by the first author at the Gutenberg
University in Mainz and by the last two authors at SISSA.
A.C. and A.T. would like to thank their supervisors,
Martin Reuter and Marco Fabbrichesi, for their continuous support.
The work of A.C. was partially supported by the Danish National Research Foundation DNRF:90 grant.
The work of A.T. was suported by the São Paulo Research Fundation (FAPESP) under grants 2011/11973-4 and 2013/02404-1.

\end{document}